\documentclass[aps,prd,preprint,groupedaddress]{revtex4}
\usepackage{amssymb,amsmath}
\usepackage{graphicx} 
\usepackage{dcolumn}  
\usepackage{feynmf}   
\usepackage{epsfig}   
\newcommand{\darrow}[1]{ \overleftrightarrow{#1} }

\newcommand{\tr}{ {\mathbf{tr}\, }}
\newcommand{\Tr}{ {\mathbf{Tr}\, }}
\renewcommand{\Re}{ {\mathbf{Re} }}

\begin{document}

\preprint{}

\title{A New Approach to Analytic, Non-Perturbative and Gauge-Invariant QCD}


\author{H. M. Fried$^{\dag}$, T. Grandou$^{\ddag}$ and Y.-M. Sheu$^{\ddag}$}
\email[]{ymsheu@alumni.brown.edu}
\affiliation{${}^{\dag}$ {Physics Department, Brown University, Providence, RI 02912, USA} \\ ${}^{\ddag}$ {Universit\'{e} de Nice Sophia-Antipolis, Institut Non Lin$\acute{e}$aire de Nice, UMR 6618 CNRS, 06560 Valbonne, France}}

\date{\today}

\begin{abstract}


Following a previous calculation of quark scattering in eikonal approximation, this paper presents a new, analytic and rigorous approach to the calculation of QCD phenomena.  In this formulation a basic distinction between the conventional "idealistic" description of QCD and a more "realistic" description is brought into focus by a non-perturbative, gauge-invariant evaluation of the Schwinger solution for the QCD generating functional in terms of the exact Fradkin representations of the Green's functional $\mathbf{G}_{c}(x,y|A)$ and the vacuum functional $\mathbf{L}[A]$. Because quarks exist asymptotically only in bound states, their transverse coordinates can never be measured with arbitrary precision; the non-perturbative neglect of this statement leads to obstructions that are easily corrected by invoking in the basic Lagrangian a probability amplitude which describes such transverse imprecision.

	The second result of this non-perturbative analysis is the appearance of a new and simplifying output called "Effective Locality", in which the interactions between quarks by the exchange of a "gluon bundle" -- which "bundle" contains an infinite number of gluons, including cubic and quartic gluon interactions -- display an exact locality property that reduces the several functional integrals of the formulation down to a set of ordinary integrals. It should be emphasized that "non-perturbative" here refers to the effective summation of all gluons between a pair of quark lines -- which may be the same quark line, as in a self-energy graph -- but does not (yet) include a summation over all closed-quark loops which are tied by gluon-bundle exchange to the rest of the "Bundle Diagram".  As an example of the power of these methods we offer as a first analytic calculation the quark-antiquark binding potential of a pion, and the corresponding three-quark binding potential of a nucleon, obtained in a simple way from relevant eikonal scattering approximations.  A second calculation, analytic, non-perturbative and gauge-invariant, of a nucleon-nucleon binding potential to form a model deuteron, will appear separately.

\end{abstract}


\maketitle

\section{\label{SEC1}Introduction}

`{\textit{The strong interactions comprise a richer field than the set of phenomena that we have learned to describe in terms of perturbative QCD or the (near-) static non-perturbative domain of lattice QCD ( ...) It may well be that interesting unusual occurrences happen outside the framework of perturbative QCD-happen in some collective, or intrinsically nonperturbative way.}}'  It is in these terms that the \textit{Resource Letter: Quantum Chromodynamics}, arXiv: 1002.5032v2 [hep-ph], of February 26, 2010, concludes its overall review of the most salient achievements realized in QCD.  Of course by its very definition, the \textit{non-perturbative} qualifier applies to a large variety of very different realizations: There are several ways of `being non-perturbative', in the either context of Lagrangian Quantum Field Theory or of Algebraic Quantum Field Theory, the former admittedly more pragmatic.

It is this route of Lagrangian Quantum Field Theory which will here be followed so as to introduce an extraordinary property of fermionic QCD amplitudes.  This property can be phrased as follows:

In any Quark/Quark (or Anti-Quark) scattering amplitude, the full
gauge-invariant sum of all cubic and quartic vectorial gluonic interactions, fermionic
loops included, results in a local contact-type interaction mediated by a tensor field which, in both internal (color) and external (Lorentz) indices is rank 2 and antisymmetric.

This property, dubbed `Effective Locality' appears to be a genuine one of QCD and is worth exploring in relation to the expected non-perturbative properties of QCD, and some examples will be presented.  It is most interesting that a previous approach, looking for a `most dual' description of QCD, and first restricted to the pure Yang-Mills case, was able to display some positive results in this direction \cite{Reinhardt:1993}. It may also be worth noting that a non-perturbative analysis of Yang-Mills thermodynamics has recently put forth the occurrence of a contact-type interaction within the confined phase of $SU(2)$ and $SU(3)$ theories \cite{Ralf:2008}. And even more recently, a fascinating functional connection of our approach to a most general formulation of quantum field theories \cite{DanielDF:2011} has been brought to our attention.

In a previous paper~\cite{Fried2009_QCD1}, it was shown how the formal Schwinger solution~\cite{HMF1,HMF2} to the QCD generating functional can, by a slight rearrangement, be brought into gauge-invariant form.  This simple procedure has been overlooked for decades. When combined with the Fradkin representations~\cite{Fradkin1966,HMF2} for the functionals $\mathbf{G}_{c}[A]$ and $\mathbf{L}[A]$ (that are an integral part of the Schwinger solution), as well as with the useful Halpern representation \cite{Halpern1977a,Halpern1977b} for $\exp{[-\frac{i}{4}\int \mathbf{F}^{2}]}$, the relevant Gaussian functional operations may be performed exactly. This corresponds to the summation of all Feynman graphs of gluons exchanged between quarks, and one then explicitly sees the cancelation of all gauge-dependent gluon propagators. Gauge invariance is achieved by gauge independence, a hope some authors long had ({\it{e.g.}} R.P. Feynman) for QED \cite{Zee:2010}, which turns out to be realized in the case of QCD thanks to its very non-abelian structure.

One also can see the appearance of a new and exact property of Effective Locality (EL), which simplifies all calculations by transforming the remaining functional integrals into sets of ordinary integrals, a non-trivial mathematical point \cite{bctg:2012}.

Further, it readily becomes apparent that one cannot continue to consider quarks in the same, conventional fashion as quanta of other ({\it{e.g.}}, abelian) fields, such as electrons, which satisfy the standard measurement properties of quantum mechanics, perfect position dependence at the cost of unknown momenta, and vice-versa.  This is impossible for quarks since they always appear asymptotically in bound states, and their transverse coordinates can never, in principle, be exactly measured.  In the developments to follow one finds that a violation of this principle produces absurdities in the exact evaluation of any QCD amplitude.  At a phenomenological level, we therefore propose a change in the basic QCD Lagrangian, which introduces a measure of transverse fluctuations by the insertion of an unknown probability amplitude that is essential to and determined from quark binding into hadrons.  All previous absurdities in estimates of all "realistic" QCD amplitudes are then removed, and one finds a simple method of constructing quark-binding potentials, as well as other, somewhat more involved potentials between nucleons to form nuclei.

This paper is divided into a sequence of Sections each emphasizing a particular aspect of this formulation.  In Section~\ref{SEC2}, as a result of the Gaussian nature of the functional operations which are performed exactly, the expression of the fundamental gauge invariance becomes evident, along with the EL property and its simplifying effects.  In Sections~\ref{SEC3},~\ref{SEC4}, and~ \ref{SEC5}, the obvious need for transverse fluctuations is made clear, and the form of a corresponding insertion of an unknown probability amplitude into the basic QCD Lagrangian is described.  Sections~\ref{SEC6},~\ref{SEC7}, and~\ref{SEC8} deal with the construction of quark-binding potentials, exhibited for a model pion and model nucleon, using a simple, analytic technique (which reverses the Potential-Theory derivation of an eikonal function from a specified potential).  Section~\ref{SEC9} is devoted to a brief Summary and Speculations.  In the remainder of the present Section, we shall for completeness summarize the argument of \cite{Fried2009_QCD1}, which shows how the conventional Schwinger solution for the generating functional may be converted, for QCD, but not for QED, into a manifestly gauge-invariant form.

It is simplest to begin with QED, and its free-photon Lagrangian,
\begin{equation}\label{Eq:1}
\mathcal{L}_{0} = - \frac{1}{4} \mathbf{f}_{\mu \nu} \mathbf{f}_{\mu \nu} =  - \frac{1}{4} \left( \partial_{\mu} A_{\nu} -  \partial_{\nu} A_{\mu} \right)^{2},
\end{equation}

\noindent whose Action Integral may be rewritten as
\begin{eqnarray}\label{Eq:2}
\int{\mathrm{d}^{4}x \, \mathcal{L}_{0}} &=&  - \frac{1}{2} \int{ \left( \partial_{\nu} A_{\mu} \right)^{2} } + \frac{1}{2} \int{ \left( \partial_{\mu} A_{\mu} \right)^{2} } \\ \nonumber &=& - \frac{1}{2} \int{ A_{\mu} \, \left( -\partial^{2}\right) \, A_{\mu} } + \frac{1}{2} \int{ \left( \partial_{\mu} A_{\mu} \right)^{2} }.
\end{eqnarray}

\noindent The difficulty of maintaining both manifest gauge invariance (MGI) and manifest Lorentz covariance (MLC) appears at this stage.  What has typically been done since the original days of Fermi, who simply neglected the inconvenient $\left( \partial_{\mu} A_{\mu} \right)^{2}$, is to use the latter to define a relativistic gauge in which all calculations maintain MLC, while relying upon strict charge conservation to maintain an effective gauge invariance of the theory.

	The choice of relativistic gauge can be arranged in various ways. The simplest functional way is to multiply the inconvenient term by the real parameter $\lambda$, and treat it as an interaction term.  For definiteness, one can begin with the free-field, ($\lambda = 0$, Feynman) propagator $\mathbf{D}_{c, \mu \nu}^{(0)} = \delta_{\mu \nu} \, \mathbf{D}_{c}$, where $(-\partial^{2}) \, \mathbf{D}_{c} = 1$; then the free-field Generating Functional (GF) is given by
\begin{equation}\label{Eq:3}
\mathfrak{Z}_{0}^{(0)}\{j\} = \exp{\left\{ \frac{i}{2} \int{j \cdot
\mathbf{D}_{\mathrm{c}}^{(0)} \cdot j} \right\} }.
\end{equation}

\noindent Then, operating upon it by the 'interaction' $\lambda$-term, a new free-field GF is produced
\begin{eqnarray}\label{Eq:4}
\mathfrak{Z}_{0}^{(\zeta)}\{j\} &=& \left. e^{\frac{i}{2} \lambda \int{ \left( \partial_{\mu} A_{\mu} \right)^{2} }} \right|_{A \rightarrow \frac{1}{i}\frac{\delta}{\delta j}} \cdot e^{\frac{i}{2} \int{j \cdot \mathbf{D}_{\mathrm{c}}^{(0)} \cdot j} } \\ \nonumber &=& e^{ \frac{i}{2} \int{j \cdot
\mathbf{D}_{\mathrm{c}}^{(\zeta)} \cdot j} } \cdot e^{-\frac{i}{2} \Tr\ln{\left[ 1 - \lambda\,( {\partial\otimes \partial/\partial^{2}}) \right]}},
\end{eqnarray}

\noindent where	$\mathbf{D}_{\mathrm{c},\mu \nu}^{(\zeta)} = \left[ \delta_{\mu \nu} - \zeta \partial_{\mu} \partial_{\nu} / \partial^{2} \right] \, \mathbf{D}_{\mathrm{c}}$, with $\zeta = \lambda/(1 - \lambda)$.  The functional operation of (\ref{Eq:4}) is fully equivalent to a bosonic, gaussian functional integration (FI). Such 'linkage operation' statements are frequently more convenient than the standard FI representations, since they do not require specification of infinite normalization constants.

	The Tr-Log term of (\ref{Eq:4}) is an infinite phase factor, representing the sum of the vacuum energies generated by longitudinal and time-like photons, with a weight $\lambda$ arbitrarily inserted; this quantity can be removed by an appropriate version of normal ordering, but can more simply be absorbed into an overall normalization constant.

	Including the conventional fermion interaction, $\mathcal{L}_{\mathrm{int}} = -ig \bar{\psi} \gamma \cdot A \psi$, and the gauge 'interaction' $\frac{1}{2} \lambda \left( \partial_{\mu} A_{\mu} \right)^{2}$, it is easy to show~\cite{HMF1,HMF2} that the standard Schwinger solution for the GF in the $\zeta$-covariant gauge becomes
\begin{eqnarray}\label{Eq:5}
\mathfrak{Z}_{\mathrm{QED}}^{(\zeta)}[j,\eta,\bar{\eta}] = \mathcal{N} \left. e^{i\int{\bar{\eta} \cdot \mathbf{G}_{\mathrm{c}}[A] \cdot \eta} + \mathbf{L}[A] + \frac{i}{2} \lambda \int{ \left( \partial_{\mu} A_{\mu} \right)^{2} }} \right|_{A \rightarrow \frac{1}{i}\frac{\delta}{\delta j}} \cdot e^{\frac{i}{2} \int{j \cdot \mathbf{D}_{\mathrm{c}}^{(0)} \cdot j} },
\end{eqnarray}

\noindent where $\mathbf{G}_{c}[A] = [m + \gamma \cdot (\partial - i g A)]^{-1}$, $\mathbf{L}[A] = \Tr\ln{\left[ 1 - ig \gamma \cdot A \mathbf{S}_{\mathrm{c}} \right]}$, $\mathbf{S}_{\mathrm{c}} = \mathbf{G}_{\mathrm{c}}[0]$, and where the phase factor of (\ref{Eq:4}) has been absorbed into $\mathcal{N}$.  A most convenient re-arrangement of (\ref{Eq:5}) uses the easily-proven identity~\cite{HMF1}, for an arbitrary functional $\mathcal{F}[A]$,
\begin{eqnarray}\label{Eq:6}
\mathcal{F}\left[ \frac{1}{i} \frac{\delta}{\delta j} \right] \cdot e^{\frac{i}{2} \int{j \cdot \mathbf{D}_{\mathrm{c}}^{(\zeta)} \cdot j} } \equiv e^{\frac{i}{2} \int{j \cdot \mathbf{D}_{\mathrm{c}}^{(\zeta)} \cdot j} } \cdot \left. e^{\mathfrak{D}_{A}} \cdot \mathcal{F}[A] \right|_{A = \int{\mathbf{D}_{\mathrm{c}}^{(\zeta)} \cdot j} },
\end{eqnarray}

\noindent where $\mathfrak{D}^{(\zeta)} _{A} =  - \frac{i}{2} \int{\frac{\delta}{\delta A} \cdot  \mathbf{D}_{\mathrm{c}}^{(\zeta)} \cdot \frac{\delta}{\delta A} }$, so that (\ref{Eq:5}) now reads
\begin{eqnarray}\label{Eq:7}
\mathfrak{Z}_{\mathrm{QED}}^{(\zeta)}[j,\eta,\bar{\eta}] = \mathcal{N} \, e^{\frac{i}{2} \int{j \cdot \mathbf{D}_{\mathrm{c}}^{(\zeta)} \cdot j} } \cdot \left. e^{\mathfrak{D}_{A}^{(\zeta)}} \cdot e^{i\int{\bar{\eta} \cdot \mathbf{G}_{\mathrm{c}}[A] \cdot \eta} + \mathbf{L}[A]} \right|_{A = \int{\mathbf{D}_{\mathrm{c}}^{(\zeta)} \cdot j}}.
\end{eqnarray}

\noindent This is the formal solution for the GF of QED in the gauge $\zeta$ that has been known and used for a half-century~\cite{HMF1}.

	We now come to QCD, with
\begin{equation}\label{Eq:8}
\mathcal{L}_{\mathrm{QCD}} = - \frac{1}{4} \mathbf{F}_{\mu \nu}^{a} \mathbf{F}_{\mu \nu}^{a} - \bar{\psi} \cdot [m + \gamma_{\mu} \, (\partial_{\mu} - i g A_{\mu}^{a} \lambda^{a})] \cdot \psi,
\end{equation}

\noindent where $\mathbf{F}_{\mu \nu}^{a} = \partial_{\mu} A_{\nu}^{a} -  \partial_{\nu} A_{\mu}^{a} + g f^{abc} A_{\mu}^{b} A_{\nu}^{c} \equiv \mathbf{f}_{\mu \nu}^{a} + g f^{abc} A_{\mu}^{b} A_{\nu}^{c}$.  One can rely on the following observations:  'proper' quantization in the Coulomb gauge, for the free and interacting theories, yields the same equal-time commutation relations (ETCRs) for QCD as for QED (with extra $\delta^{ab}$ color factors appearing in all relevant equations); at $g = 0$, QCD is the same free-field theory as QED (except for additional color indices); QED in any of the conventional relativistic gauges can be obtained by treating the $\frac{1}{2} \lambda \left( \partial_{\mu} A_{\mu} \right)^{2}$ as an 'interaction' (as above). Taking advantage of these observations, one can set up QCD in the form used above for QED.

	As a final preliminary step, we write
\begin{equation}\label{Eq:9}
-\frac{1}{4} \int{\mathbf{F}^{2}} = -\frac{1}{4} \int{\mathbf{f}^{2}} - \frac{1}{4} \int{\left[\mathbf{F}^{2} - \mathbf{f}^{2} \right]} \equiv -\frac{1}{4} \int{\mathbf{f}^{2}} + \int{\mathcal{L}'[A]},
\end{equation}

\noindent with $\mathbf{f}_{\mu \nu}^{a} = \partial_{\mu} A_{\nu}^{a} -  \partial_{\nu} A_{\mu}^{a}$ and $\mathcal{L}'[A] =  -\frac{1}{4} \left(2 \mathbf{f}_{\mu \nu}^{a} +  g f^{abc} A_{\mu}^{b} A_{\nu}^{c} \right) \, \left(g f^{abc} A_{\mu}^{b} A_{\nu}^{c} \right)$, and for subsequent usage, after an integration-by-parts, we note the exact relation
\begin{equation}\label{Eq:10}
-\frac{1}{4} \int{\mathbf{F}^{2}} = - \frac{1}{2} \int{ A_{\mu}^{a} \, \left( -\partial^{2}\right) \, A_{\mu}^{a} } + \frac{1}{2} \int{ \left( \partial_{\mu} A_{\mu}^{a} \right)^{2} } + \int{\mathcal{L}'[A]},
\end{equation}

\noindent (In the next few paragraphs, for simplicity, we suppress the fermion/quark variables, which will be re-inserted at the end of this discussion).

	In order to select a particular relativistic gauge, one can multiply the 2nd RHS term of (\ref{Eq:10}) by $\lambda$, and include this term as part of the interaction, thereby obtaining the familiar QCD GF in the relativistic gauge specified by
\begin{eqnarray}\label{Eq:11}
\mathfrak{Z}_{\mathrm{QCD}}^{(\zeta)}[j] = \mathcal{N} \, e^{i \int{\mathcal{L}'\left[\frac{1}{i} \, \frac{\delta}{\delta j} \right]}} \cdot e^{\frac{i}{2} \lambda \int{\frac{\delta}{\delta j_{\mu}} \, \partial_{\mu} \partial_{\nu} \, \frac{\delta}{\delta j_{\nu}}} } \cdot e^{\frac{i}{2} \int{j \cdot \mathbf{D}_{\mathrm{c}}^{(0)} \cdot j} },
\end{eqnarray}

\noindent or, after re-arrangement,
\begin{eqnarray}\label{Eq:12}
\mathfrak{Z}_{\mathrm{QCD}}^{(\zeta)}[j] = \mathcal{N} \, e^{i \int{\mathcal{L}'\left[\frac{1}{i} \, \frac{\delta}{\delta j} \right]}} \cdot e^{\frac{i}{2} \int{j \cdot \mathbf{D}_{\mathrm{c}}^{(\zeta)} \cdot j} }
\end{eqnarray}

\noindent with the determinantal phase factor of (\ref{Eq:4}) included in the normalization $\mathcal{N}$, and a $\delta^{ab}$ associated with each free-gluon propagator $\mathbf{D}_{\mathrm{c}}$.

	After re-inserting the quark variables, and after re-arrangement, expansion of (\ref{Eq:12}) in powers of $g$ clearly generates the conventional Feynman graphs of perturbation theory in the gauge $\zeta$.  It is clear that all choices of $\lambda$ are possible except $\lambda = 1$, for that choice leads to $\zeta \rightarrow \infty$ and an ill-defined gluon propagator.  This is an unfortunate situation, because the choice $\lambda = 1$ is precisely the one that corresponds to MGI in QCD, as is clear from (\ref{Eq:10}).

	But there is a very simple way of re-writing (\ref{Eq:12}), by replacing the $\int{\mathcal{L}'[A]}$ of that equation by the relation given by (\ref{Eq:10}),
\begin{equation}\label{Eq:13}
i \int{\mathcal{L}'[A]} = -\frac{i}{4} \int{\mathbf{F}^{2}} + \frac{i}{2} \int{ A_{\mu}^{a} \, \left( -\partial^{2}\right) \, A_{\mu}^{a} } -\frac{i}{2} \int{ \left( \partial_{\mu} A_{\mu}^{a} \right)^{2} },
\end{equation}

\noindent which (continuing to suppress the quark variables momentarily) yields
\begin{eqnarray}\label{Eq:14}
\mathfrak{Z}_{\mathrm{QCD}}^{(\zeta)}[j] = \mathcal{N} \, \left. e^{-\frac{i}{4} \int{\mathbf{F}^{2}} - \frac{i}{2} (1 - \lambda) \int{ \left( \partial_{\mu} A_{\mu}^{a} \right)^{2}} + \frac{i}{2} \int{ A_{\mu}^{a} \, \left( -\partial^{2}\right) \, A_{\mu}^{a} } } \right|_{A \rightarrow \frac{1}{i} \, \frac{\delta}{\delta j} } \cdot e^{\frac{i}{2} \int{j \cdot \mathbf{D}_{\mathrm{c}}^{(0)} \cdot j} }.
\end{eqnarray}

It is now obvious that the choice $\lambda =1$ can be made. It will become clear below that, using (\ref{Eq:14}), the functional operations effectively treat gluons as if they were quanta of a 'ghost' field.  The gluons of the theory, never measurable by themselves, disappear effectively from the exact calculation of every QCD correlation function, without approximation and without exception.  This 'ghost mechanism' occurs because all factors of
\begin{equation*}
e^{\frac{i}{2} \int{j \cdot \mathbf{D}_{\mathrm{c}}^{(0)} \cdot j} }
\end{equation*}

\noindent of (\ref{Eq:14}) are, in the sum of all virtual gluon processes, effectively removed by the action of the term
\begin{equation*}
\left. e^{\frac{i}{2} \int{ A_{\mu}^{a} \, \left( -\partial^{2}\right) \, A_{\mu}^{a} } }\right|_{A \rightarrow \frac{1}{i} \, \frac{\delta}{\delta j} }
\end{equation*}

\noindent of (\ref{Eq:14}).  In the end, the gluon acts as a 'spark plug' to generate the MGI and MLC interactions of the theory, which then take on remarkably simple forms.		

	However, if one is interested in the radiative corrections to the gauge-dependent, free-field gluon propagator, which corrections are now guaranteed to be gauge-independent, the leading RHS factor of
\begin{equation*}
e^{\frac{i}{2} \int{j \cdot \mathbf{D}_{\mathrm{c}}^{(0)} \cdot j} }
\end{equation*}

\noindent should be retained in the rearranged expression of (\ref{Eq:14}), taken at $\lambda = 1$,
\begin{eqnarray}\label{Eq:15}
\mathfrak{Z}_{\mathrm{QCD}}^{(0)}[j] = {\mathcal{N}}e^{\frac{i}{2} \int{j \cdot \mathbf{D}_{\mathrm{c}}^{(0)} \cdot j} } \cdot \left. e^{- \frac{i}{2} \int{\frac{\delta}{\delta A} \cdot \mathbf{D}_{\mathrm{c}}^{(0)} \cdot \frac{\delta}{\delta A} } } \cdot e^{-\frac{i}{4} \int{\mathbf{F}^{2}} + \frac{i}{2} \int{ A \cdot \left( -\partial^{2}\right) \cdot A} } \right|_{A = \int{\mathbf{D}_{\mathrm{c}}^{(0)} \cdot j} }.
\end{eqnarray}

\noindent Otherwise, and for the specific examples to follow, this factor plays no role and will be suppressed. The resulting GF is then MGI, and its superscript will be suppressed.

	After re-inserting quark variables, (\ref{Eq:15}) becomes
\begin{eqnarray}\label{Eq:16}
\mathfrak{Z}_{\mathrm{QCD}}[j, \bar{\eta}, \eta] = \mathcal{N} \, \left. e^{- \frac{i}{2} \int{\frac{\delta}{\delta A} \cdot \mathbf{D}_{\mathrm{c}}^{(0)} \cdot \frac{\delta}{\delta A} } } \cdot e^{-\frac{i}{4} \int{\mathbf{F}^{2}} + \frac{i}{2} \int{ A \cdot \left( -\partial^{2}\right) \cdot A} } \cdot e^{i\int{\bar{\eta} \cdot \mathbf{G}_{\mathrm{c}}[A] \cdot \eta} + \mathbf{L}[A]}\right|_{A = \int{\mathbf{D}_{\mathrm{c}}^{(0)} \cdot j} },
\end{eqnarray}

\noindent and we next invoke the representation suggested by Halpern~\cite{Halpern1977a,Halpern1977b}
\begin{equation}\label{Eq:17}
e^{-\frac{i}{4} \int{\mathbf{F}^{2}}} = \mathcal{N}' \, \int{\mathrm{d}[\chi] \, e^{ \frac{i}{4} \int{ \left(\chi_{\mu \nu}^{a}\right)^{2} + \frac{i}{2} \int{ \chi^{\mu \nu}_{a} \mathbf{F}_{\mu \nu}^{a}} } } },
\end{equation}

\noindent where
\begin{equation}
\int{\mathrm{d}[\chi]} = \prod_{i} \prod_{a} \prod_{\mu \nu} \int{\mathrm{d}\chi_{\mu \nu}^{a}(w_{i})},
\end{equation}

\noindent so that (\ref{Eq:17}) represents a functional integral over the anti-symmetric tensor field $\chi_{\mu \nu}^{a}$.  Following the standard definition~\cite{Zee:2010}, all space-time is broken up into small cells of size $\delta^{4}$ about each point $w_{i}$, and $\mathcal{N}'$ is a normalization constant so chosen such that the right hand side of (\ref{Eq:17}) becomes equal to unity as $\mathbf{F}_{\mu \nu}^{a} \rightarrow 0$.  In this way, the GF may be re-written as ($\mathcal{N}' \cdot \mathcal{N} = \mathcal{N}''\rightarrow \mathcal{N}$)
\begin{eqnarray}\label{Eq:18}
\mathfrak{Z}_{\mathrm{QCD}}[j, \bar{\eta}, \eta] = \mathcal{N} \, \int{\mathrm{d}[\chi] \, e^{ \frac{i}{4} \int{ \chi^{2} }} } \, \left. e^{\mathfrak{D}_{A}^{(0)}} \cdot e^{+\frac{i}{2} \int{\chi \cdot \mathbf{F} + \frac{i}{2} \int{ A \cdot \left( -\partial^{2}\right) \cdot A} }} \cdot e^{i\int{\bar{\eta} \cdot \mathbf{G}_{\mathrm{c}}[A] \cdot \eta} + \mathbf{L}[A]}\right|_{A = \int{\mathbf{D}_{\mathrm{c}}^{(0)} \cdot j} },
\end{eqnarray}

\noindent where $\exp{[\mathfrak{D}_{A}^{(0)}]}$ is the linkage operator with $\mathfrak{D}_{A}^{(0)} = - \frac{i}{2} \int{\frac{\delta}{\delta A} \cdot \mathbf{D}_{\mathrm{c}}^{(0)} \cdot \frac{\delta}{\delta A}}$.

	Quarks and anti-quarks are treated as stable entities during any scattering, production or binding process, which means that relevant functional derivatives with respect to the sources $\eta$, $\bar{\eta}$, will bring down factors of $\mathbf{G}_{\mathrm{c}}(x,y|A)$, one such factor for each quark or anti-quark under discussion.  For example, by standard mass-shell amputation, one can then pass to the construction of a scattering amplitude of a pair of quarks, or of any number of quarks, or of a quark-anti-quark pair. If this scattering is to occur at high-energies and small momentum transfer, a convenient and relatively simple eikonal approximation is available, derived in detail in Appendix B of Ref.~\cite{Fried2000}.  This will be a most convenient tool in Section~\ref{SEC7}, where we employ the well-known connection between an eikonal function dependent upon impact parameter, and an effective potential which is the cause of the scattering or production or binding which leads to that eikonal function.

	But it is worth emphasizing that any such, simplifying eikonal approximation is not to be confused with the exact functional representations corresponding to scattering, production and binding.  Because the linkage operator in (\ref{Eq:18}) represents an effective Gaussian functional operation upon the $A$-dependence contained within $\mathbf{G}_{\mathrm{c}}[A]$ and $\mathbf{L}[A]$, and because there exist Fradkin representations~\cite{Fradkin1966} of these functionals which are Gaussian in $A$, the specific functional operations required, resulting from well-defined functional derivatives with respect to the sources, $\eta$, $\bar{\eta}$, according to the physical process under consideration, can be performed exactly.  This produces the sum of all Feynman graphs corresponding to the exchange of an infinite number of gluons between quarks and/or anti-quarks, exhibited in terms of the Fradkin functional parameters that define the representations for $\mathbf{G}_{\mathrm{c}}[A]$ and $\mathbf{L}[A]$. For clarity and convenience, we reproduce exact expressions of these Fradkin representations in Appendix~\ref{AppA}.

	The result of this approach is gauge-invariant and contains new structures and requirements of a non-perturbative nature. It displays results of such simplicity that one is able, as in Section~\ref{SEC7}, to present an analytic derivation of quark binding potentials. A separate paper using this approach will present an analytic derivation of a nucleon-nucleon scattering and (deuteron) binding potential, which is, to our knowledge, the very first analytic example of Nuclear Physics derived from basic QCD.

\section{\label{SEC2}Explicit Gauge Invariance}

	The correlation functions of QCD are obtained by appropriate functional differentiation of (\ref{Eq:18}) with respect to gluon and quark sources. Since we are here concerned only with quark ($Q$) or anti-quark ($\bar{Q}$) interactions, in which all possible numbers of virtual gluons are exchanged, we immediately set the gluon sources $j$ equal to zero.  All $Q$/$\bar{Q}$ amplitudes are then obtained by pairwise functional differentiation of the quark sources $\eta(y)$, $\bar{\eta}(x)$, and each such operation "brings down" one of a set of (properly anti-symmetrized) Green's functions, $\mathbf{G}_{\mathrm{c}}[A]$.  For example, the 2-point quark propagator will involve the FI $\int{d[\chi]}$ and the linkage operator acting upon $\mathbf{G}_{\mathrm{c}}(x,y|A) \, \exp{\{\mathbf{L}[A]\}}$, followed by setting $A \rightarrow 0$.  Similarly, the $Q$/$\bar{Q}$ scattering amplitude will be obtained from the same functional operations acting upon the (anti-symmetrized) combination $\mathbf{G}_{\mathrm{c}}[A] \mathbf{G}_{\mathrm{c}}[A] \exp{\{\mathbf{L}[A]\}}$, followed by $A \rightarrow 0$, as
\begin{eqnarray}\label{Eq:19}
& & \mathbf{M}(x_{1}, y_{1}; x_{2}, y_{2}) \\ \nonumber  & & = \left. \frac{\delta}{\delta \bar{\eta}(y_{1})} \cdot \frac{\delta}{\delta \eta(x_{1})} \cdot \frac{\delta}{\delta \bar{\eta}(y_{2})} \cdot \frac{\delta}{\delta \eta(x_{2})} \cdot \mathfrak{Z}_{c}\left\{ j, \bar{\eta}, \eta \right\} \right|_{\eta=\bar{\eta}=0; j=0} \\ \nonumber & & = \, \mathcal{N} \, \int{d[\chi] \, e^{\frac{i}{4}\int{\chi^{2}}} \, e^{\mathfrak{D}_{A}^{(0)}} \,} e^{ + \frac{i}{2}\int{\chi\cdot \mathbf{F}} +\frac{i}{2}\int{A \cdot \left(\mathbf{D}_{c}^{(0)}\right)^{-1} \cdot A }} \, \left. \mathbf{G}_{\mathrm{c}}(x_{1}, y_{1}|gA) \, \mathbf{G}_{\mathrm{c}}(x_{2}, y_{2}|gA) \, e^{\mathbf{L}[A]} \right|_{A=0} \\ \nonumber & & \quad \quad \quad - \{ {1} \leftrightarrow {2} \},
\end{eqnarray}

\noindent and other fermionic 2n-point functions are obtained in the same way.

	The Fradkin functional representations for $\mathbf{G}_{\mathrm{c}}[A]$ and $\mathbf{L}[A]$, derived in Appendix~\ref{AppA}, display a Gaussian dependence on $A$, and hence the linkage operations of (\ref{Eq:19}), in any order of the expansion of $\exp\{\mathbf{L}[A]\}$ in powers of $\mathbf{L}$, or somewhat more conveniently, using a functional cluster expansion~\cite{HMF1} for $\exp\{\mathfrak{D}_{A}^{(0)}\}$ operating upon $\exp\{\mathbf{L}[A]\}$, can be performed exactly.  One small complication, easily surmounted, is due to the non-Abelian nature of QCD, in which the $A$-dependence appears inside an ordered exponential, ordered in terms of an invariant "Schwinger proper time" variable, $s$, as in the explicit representation for $\mathbf{G}_{\mathrm{c}}(x,y|A)$ and $\mathbf{L}[A]$.  However cumbersome these representations may appear, the essential fact is that they are Gaussian in $A$, and hence the linkage operations, corresponding to the summation over all possible gluon exchanges between $Q$ and/or $\bar{Q}$ lines, may be performed exactly.  One then finds a simple, effectively local representation for the sum over all such exchanges, here called a "gluon bundle". Henceforth, "bundle diagrams" will replace Feynman diagrams containing individually-specified gluon exchanges.

	To display the gauge invariance of all such QCD correlation functions, one first notes that $\mathbf{L}[A]$ is explicitly invariant under arbitrary changes of the full QCD gauge group~\cite{HMF2}.  Then, it is convenient to combine the Gaussian $A$-dependence of every entering $\mathbf{G}_{\mathrm{c}}[A]$ into the quantity
\begin{eqnarray}\label{Eq:20}
\exp{\left[ \frac{i}{2}\int{\mathrm{d}^{4}z \, A^{\mu}_{a}(z) \, \tilde{\mathbf{K}}_{\mu \nu}^{ab}(z) \, A^{\nu}_{b}(z)} + i\int{\mathrm{d}^{4}z \, \tilde{\mathbf{Q}}^{\mu}_{a}(z) \, A_{\mu}^{a}(z)} \right]},
\end{eqnarray}

\noindent where $\tilde{\mathbf{K}}$ and $\tilde{\mathbf{Q}}$ are local functions of the Fradkin variables, collectively denoted by $u_{\mu}(s')$, and the $\Omega^{a}(s_{1})$, $\Omega^{a}(s_{2})$, $\Phi_{\mu \nu}^{a}(s_{1})$ and $\Phi_{\mu \nu}^{a}(s_{2})$ are needed to extract the $A_{\mu}^{a}(y - u(s'))$ from ordered exponentials.  Note that the $\tilde{\mathbf{Q}}$ and $\tilde{\mathbf{K}}$ are also to represent the sum of similar contributions from each of the $\mathbf{G}_{\mathrm{c}}(x,y|A)$ which collectively generate the amplitude under consideration.  For example, in the case of the 4-point function one will obtain from the product of $\mathbf{G}_{\mathrm{c}, \mathrm{I}}[A]$ and $\mathbf{G}_{\mathrm{c}, \mathrm{I\!I}}[A]$,
\begin{eqnarray}\label{Eq:21}
\tilde{\mathbf{K}}_{\mu \nu}^{ab}(z) = & & 2 g^{2} \int_{0}^{s_{1}}{\mathrm{d}s' \, \delta^{(4)}(z - y_{1} + u(s')) f^{abc} \Phi_{\mu\nu,\mathrm{I}}^{c}(s')} \\ \nonumber & & + 2 g^{2} \int_{0}^{s_{2}}{\mathrm{d}s' \, \delta^{(4)}(z - y_{2} + \bar{u}(s')) f^{abc} \Phi_{\mu\nu,\mathrm{I\!I}}^{c}(s')},
\end{eqnarray}

\noindent and
\begin{eqnarray}\label{Eq:22}
\tilde{\mathbf{Q}}_{\mu \nu}^{a}(z) = & & - 2g \partial_{\nu} \Phi_{\nu \mu, \mathrm{I}}^{a}(z) - g \int_{0}^{s_{1}}{\mathrm{d}s' \, \delta^{(4)}(z - y_{1} +u(s')) \, u'_{\mu}(s') \Omega_{\mu,\mathrm{I}}^{a}(s')} \\ \nonumber & & - 2g \partial_{\nu} \Phi_{\nu \mu, \mathrm{I\!I}}^{a}(z) - g \int_{0}^{s_{2}}{\mathrm{d}s' \, \delta^{(4)}(z - y_{2} + \bar{u}(s')) \, \bar{u}'_{\mu}(s') \Omega_{\mu,\mathrm{I\!I}}^{a}(s')},
\end{eqnarray}

\noindent where subscripts 1,2 and $\mathrm{I}$, $\mathrm{I\!I}$ are used interchangeably to denote particles 1 and 2; and, for clarity, $\bar{u}(s')$ is used to denote particle 2, and the function
\begin{eqnarray}\label{Eq:23}
\Phi_{\mu \nu}^{a}(z) \equiv \int_{0}^{s}{\mathrm{d}s' \, \delta^{(4)}(z - y +u(s')) \, \Phi_{\mu \nu}^{a}(s')}
\end{eqnarray}

\noindent is introduced in $\tilde{\mathbf{Q}}$ for ease of presentation.  For higher quark n-point functions, there will be additional terms contributing to $\tilde{\mathbf{Q}}$ and $\tilde{\mathbf{K}}$, but their forms will be the same.

	Combining the quadratic and linear $A$-dependence from $\tilde{\mathbf{K}}$ and $\tilde{\mathbf{Q}}$ above, and including that $\mathbf{L}[A]$ dependence explicitly written in (\ref{Eq:18}), the needed linkage operation reads
\begin{eqnarray}\label{Eq:24}
\exp{\left[- \frac{i}{2} \int{\frac{\delta}{\delta A} \cdot \mathbf{D}_{\mathrm{c}}^{(0)} \cdot \frac{\delta}{\delta A}}\right]} \cdot \exp{\left[ \frac{i}{2}\int{A \cdot \bar{\mathbf{K}} \cdot A} + i \int{\bar{\mathbf{Q}} \cdot  A} \right]} \cdot \exp{\left( \mathbf{L}[A] \right)},
\end{eqnarray}

\noindent where
\begin{eqnarray}\label{Eq:25}
\langle z | \bar{\mathbf{K}}_{\mu \nu}^{ab} |z' \rangle = \left[ \tilde{\mathbf{K}}_{\mu \nu}^{ab}(z) + g f^{abc} \chi_{\mu \nu}^{c}(z) \right] \, \delta^{(4)}(z - z') + \langle z | \left. \left(\mathbf{D}_{c}^{(0)} \right)^{-1} \right|_{\mu \nu}^{ab} |z' \rangle
\end{eqnarray}

\noindent and
\begin{eqnarray}\label{Eq:26}
\bar{\mathbf{Q}}_{\mu}^{a}(z) = \tilde{\mathbf{Q}}_{\mu}^{a}(z) + \partial_{\nu} \chi_{\nu \mu}^{a}(z).
\end{eqnarray}

\noindent In $\bar{\mathbf{K}}$, all terms but the inverse of the gluon propagator are local.

Eq.~(\ref{Eq:24}) requires the linkage operator to act upon the product of two functionals of $A$, and this can be represented by the identity
\begin{eqnarray}\label{Eq:27}
e^{\mathfrak{D}_{A}} \, \mathcal{F}_{1}[A] \mathcal{F}_{2}[A] = \left( e^{\mathfrak{D}_{A}} \, \mathcal{F}_{1}[A] \right) \,  e^{\darrow{\mathfrak{D}}} \, \left( e^{\mathfrak{D}_{A}} \, \mathcal{F}_{2}[A] \right),
\end{eqnarray}

\noindent where, with an obvious notation, the 'cross-linkage' operator $\exp\{\darrow{\mathfrak{D}}\}$ is defined by
\begin{equation}\label{Eq:28}
\overleftrightarrow{\mathfrak{D}} = -i \int{\overleftarrow{\frac{\delta}{\delta A}}  \mathbf{D}_{\mathrm{c}}^{(0)} \overrightarrow{\frac{\delta}{\delta A}}}.
\end{equation}

\noindent With the identifications
\begin{eqnarray}\label{Eq:29}
\mathcal{F}_{1}[A] =  \exp{\left[ \frac{i}{2}\int{A \cdot \bar{\mathbf{K}} \cdot A} + i \int{\bar{\mathbf{Q}} \cdot  A} \right]}, \quad   \mathcal{F}_{2}[A] = \exp{\left( \mathbf{L}[A] \right)},
\end{eqnarray}

\noindent the evaluation of $e^{\mathfrak{D}_{A}} \, \mathcal{F}_{1}[A]$ is given by a standard, functional identity~\cite{HMF1,HMF2}, and reads
\begin{eqnarray}\label{Eq:30}
e^{\mathfrak{D}_{A}} \, \mathcal{F}_{1}[A] &=& \exp{\left[ \frac{i}{2} \int{ \bar{\mathbf{Q}} \cdot \mathbf{D}_{c}^{(0)} \cdot (1 - \bar{\mathbf{K}} \cdot \mathbf{D}_{c}^{(0)})^{-1} \cdot \bar{\mathbf{Q}}} -\frac{1}{2} \Tr{\ln{\left(1-\mathbf{D}_{c} \cdot \bar{\mathbf{K}}\right)}} \right]} \\ \nonumber & & \quad \cdot \exp{\left[ \frac{i}{2} \int{ A \cdot \bar{\mathbf{K}} \cdot \left( 1 - \mathbf{D}_{c}^{(0)} \cdot \bar{\mathbf{K}} \right)^{-1} \cdot A} + i \int{\bar{\mathbf{Q}} \cdot \left( 1- \bar{\mathbf{K}} \cdot \mathbf{D}_{c}^{(0)} \right)^{-1} \cdot A} \right]}.
\end{eqnarray}

\noindent It is to be noted that the kernel $\mathbf{D}_{c}^{(0)} \cdot \left( 1 - \bar{\mathbf{K}} \cdot \mathbf{D}_{c}^{(0)} \right)^{-1}  $ reduces to
\begin{equation}\label{Eq:31}
\mathbf{D}_{c}^{(0)} \cdot \left( 1 - \bar{\mathbf{K}} \cdot \mathbf{D}_{c}^{(0)} \right)^{-1} = \mathbf{D}_{c}^{(0)}  \cdot \left[ 1 -  (\widehat{\mathbf{K}}  + \left(\mathbf{D}_{c}^{(0)}\right) ^{-1}) \cdot \mathbf{D}_{c}^{(0)} \right]^{-1} = - {\widehat{\mathbf{K}}}^{-1},
\end{equation}

\noindent where instead of (\ref{Eq:25}), one now has
\begin{eqnarray}\label{Eq:32}
\widehat{\mathbf{K}}_{\mu \nu}^{ab} = \tilde{\mathbf{K}}_{\mu \nu}^{ab} + g f^{abc} \chi_{\mu \nu}^{c}.
\end{eqnarray}

\noindent It is interesting to note that the additive character of (\ref{Eq:32}) in both spin and 'isospin'-related tensors appears as a natural structure of this formulation~\cite{Huang:1977}.

	In the limit $A \rightarrow 0$, Eq.~(\ref{Eq:27}) now yields
\begin{eqnarray}\label{Eq:33}
& & e^{\mathfrak{D}_{A}} \, \mathcal{F}_{1}[A] \mathcal{F}_{2}[A] \\ \nonumber &=& \exp{\left[ -\frac{i}{2} \int{\bar{\mathbf{Q}} \cdot \widehat{\mathbf{K}}^{-1} \cdot \bar{\mathbf{Q}}} + \frac{1}{2} \Tr\ln{\widehat{\mathbf{K}}} +  \frac{1}{2} \Tr\ln{\left(-\mathbf{D}_{\mathrm{c}}^{(0)}\right)} \right]} \\ \nonumber & & \cdot \exp{\left[+ \frac{i}{2} \int{\frac{\delta}{\delta A'} \cdot \mathbf{D}_{\mathrm{c}}^{(0)} \cdot \frac{\delta}{\delta A'}}\right]} \cdot \exp{\left[ \frac{i}{2} \int{\frac{\delta}{\delta A'} \cdot \widehat{\mathbf{K}}^{-1} \cdot \frac{\delta}{\delta A'}} - \int{\bar{\mathbf{Q}} \cdot \widehat{\mathbf{K}}^{-1} \cdot \frac{\delta}{\delta A'} }\right]} \\ \nonumber & & \cdot \left( e^{\mathfrak{D}_{A'}} \, \mathcal{F}_{2}[A'] \right).
\end{eqnarray}

\noindent Now observe that the first exponential term on the second line of (\ref{Eq:33}) is exactly $\exp{\{- \mathfrak{D}_{A'}\}}$, and serves to remove the $\exp{\{\mathfrak{D}_{A'}\}}$ of the operation $\left(\exp{\{\mathfrak{D}_{A'}\}} \cdot \mathcal{F}_{2}[A']\right)$.  With the exception of an irrelevant $\exp{\left[\Tr\ln{\left(-\mathbf{D}_{\mathrm{c}}^{(0)}\right)}\right]}$ factor, to be absorbed into an overall normalization, what remains to all orders of coupling for every such process is therefore the generic structure
\begin{eqnarray}\label{Eq:34}
e^{\mathfrak{D}_{A}} \, \mathcal{F}_{1}[A] \mathcal{F}_{2}[A] = & & \mathcal{N} \, \exp{\left[ -\frac{i}{2} \int{\bar{\mathbf{Q}} \cdot \widehat{\mathbf{K}}^{-1} \cdot \bar{\mathbf{Q}}} + \frac{1}{2} \Tr\ln{\widehat{\mathbf{K}}} \right]} \\ \nonumber & & \cdot \exp{\left[ \frac{i}{2} \int{\frac{\delta}{\delta A} \cdot \widehat{\mathbf{K}}^{-1} \cdot \frac{\delta}{\delta A}} - \int{\bar{\mathbf{Q}} \cdot \widehat{\mathbf{K}}^{-1} \cdot \frac{\delta}{\delta A} }\right]} \cdot \exp{\left( \mathbf{L}[A] \right)},
\end{eqnarray}

\noindent in which the now-useless prime of $A'$ has been suppressed.  From (\ref{Eq:34}) one may now draw the following conclusions:

	(i) Nothing in Eq.~(\ref{Eq:34}) refers to $\mathbf{D}_{\mathrm{c}}^{(0)}$, which means that gauge invariance is here rigorously achieved as a matter of gauge independence.  Such invariance cannot be more manifest. The importance and novelty of this exact formulation can be recognized and appreciated in view, for example, of Ref.~\cite{Zee:2010} and \cite{Guay:2004}.

	(ii) As expressed in Eq.~(\ref{Eq:34}), each linkage operation upon $\exp{\{\mathbf{L}[A]\}}$ now consists of the exchange of a full 'bundle' of gluons, represented by a factor of $\widehat{\mathbf{K}}^{-1}$, while all of the cubic and quartic gluon interactions are conveniently incorporated into the remaining Halpern functional integral.  This is quite different, in structure and interpretation, from linkages involving the corresponding $\mathbf{L}[A]$ of an abelian theory, which entail the exchange of but a single virtual boson.  It now becomes convenient to replace Feynman diagrams, containing different orders of gluon exchange, by 'bundle diagrams', in which every graph depicts the non-perturbative exchange of all possible gluons between a pair of quark and/or anti-quark lines.  This will be described, in detail, in the next Section.

	It should be noted that quite similar forms involving at least the part $(g f \cdot \chi)^{-1}$ of $\widehat{\mathbf{K}}^{-1}$, were previously obtained in an instanton approximation to a gauge-dependent functional integral over gluon fluctuations \cite{Reinhardt:1993}.  The present result, Eq.~(\ref{Eq:34}), shows that such forms are an integral part of the exact QCD theory.  This manifest gauge-invariant construction does not work for QED, where the simple rearrangement leading from Eq.~(\ref{Eq:12}) to (\ref{Eq:14}) cannot be implemented.

	(iii) A striking aspect of Eq.~(\ref{Eq:34}) is that, because $\widehat{\mathbf{K}} = \tilde{\mathbf{K}} + (g f \cdot \chi)$ and the $\tilde{\mathbf{K}}$ and $\tilde{\mathbf{Q}}$ coming from $\mathbf{L}[A]$ are all local functions, with non-zero matrix elements $\langle z | \tilde{\mathbf{K}} | z' \rangle = \tilde{\mathbf{K}}(z) \, \delta^{(4)}(z - z')$, the contributions of Eq.~(\ref{Eq:34}) will depend on the Fradkin and Halpern variables in a specific but local way.  This remarkable property will hereafter be called 'Effective Locality' (EL), and one can now display the practical usefulness of this description in which relevant functional integrals can effectively be reduced to a few sets of ordinary integrals \cite{bctg:2012}.

\section{\label{SEC3}EFFECTIVE LOCALITY}

It is worth emphasizing the locality aspects of the results in the previous Section. Perhaps, the simplest example of EL is in the context of the simple, but non-trivial, quenched eikonal scattering model studied in Ref.~\cite{Fried2009_QCD1}, which contained the Halpern functional integral
\begin{eqnarray}\label{Eq:35}
\mathcal{N} \, \int{\mathrm{d}[\chi] \, e^{\frac{i}{4} \, \int{\chi^{2}} } \cdot \left[ \det{\left( g f \cdot \chi \right)}\right]^{-\frac{1}{2}} \cdot \exp{\left[ -\frac{i}{2} \int{\bar{\mathbf{Q}} \cdot \left( g f \cdot \chi \right)^{-1} \cdot \bar{\mathbf{Q}}} \right]}},
\end{eqnarray}

\noindent where the neglect of terms contributing to possible quark self-energies, in the limit of strong coupling, replace the exponential factor of (\ref{Eq:35}) by the argument
\begin{equation}\label{Eq:36}
ig \, \varphi(b) \, \Omega_{\mathrm{I}}^{a} \, \left. \left[ f \cdot \chi(w)\right]^{-1} \right|_{03}^{ab} \, \Omega_{\mathrm{I\!I}}^{a}.
\end{equation}

\noindent The Fradkin $u'$-variables have been replaced by asymptotic 4-momenta by virtue of the simplifying eikonal approximation, and those momenta are automatically cancelled as will be made clear after (\ref{Eq:typic}) and (\ref{Eq:41'}).  In (\ref{Eq:36}), the color factors $\Omega_{\mathrm{I}}^{a}(0)$, $\Omega_{\mathrm{I\!I}}^{b}(0)$ are peaked at $s_1=0$ and $s_2=0$, as will be noted shortly in conjunction with the function $\varphi(b)$, which depends on the collision's impact parameter $b$.

	In Ref.~\cite{Fried2009_QCD1}, as a result of EL, the argument of the Halpern variable $\chi(w)$ of (\ref{Eq:36}) was shown to be fixed at a specific value $w_{0} = (0, \vec{y}_{\perp},0)$~\cite{Note}, where $y$ denotes the CM space-time coordinate of one of the scattering quark or antiquark.  This is the only $\chi(w)$ that is relevant to the interaction. All of the other $\chi(w)$, for $w \neq w_{0}$ (and surrounded by a small volume of amount $\delta^{4}$, as in the definition/construction of a functional integral~\cite{Zee:2010}) are simply removed from the problem along with their normalization factors, leaving a single, normalized functional integral over $\mathrm{d}^{n}\chi(w_{0})$.
\par
  Further, were the values of $y$ to be subsequently changed, so that $w_{0} \rightarrow w_{1}$, then all the $\chi(w)$, $w \neq w_{1}$, would become irrelevant, and with their normalization factors will cancel away.  In effect, the space-time index $w$ is deprived of any physical meaning, and in this simple, quenched eikonal scattering, can be omitted  [When quenching is removed and different factors of $\mathbf{L}[A]$ are introduced, there will be more than one such $\int{\mathrm{d}^{n}\chi(w)}$ to be performed; but the essential, simplifying effect of EL remains].  In brief, thanks to the EL property of this formulation, the Halpern functional integral can be reduced to a set of ordinary finite dimensional integrals~\cite{bctg:2012}, which can be evaluated numerically, or approximated by a relevant physical approximation.



%
%
%
\section{\label{SEC4}QCD transverse fluctuations}

A basic distinction between QCD and other theories, in particular QED, must now be made, following from the materials of Sec.~\ref{SEC2} and \ref{SEC3}: It is central to all applications to follow.

That distinction occurs because the quanta of all (abelian) quantized fields may be expected to obey standard quantum-mechanical properties, such as perfect position dependence at the cost of unknown momenta, and vice-versa.  But this is impossible for quarks since they always appear asymptotically in bound states. Their transverse positions or momenta can never, in principle, be exactly measured.  Neglect of this distinction produces an absurdity in the exact evaluation of all QCD amplitudes.

A phenomenological change in the basic QCD Lagrangian will accordingly be proposed, such that a probability amplitude of transverse fluctuations is automatically contained in the new Lagrangian, which eventually leads to those potentials essential to quark binding into hadrons, and hadron binding into nuclei.  Then, all absurdities in estimates of QCD amplitudes are removed, enabling one to analytically calculate the effective potentials that produce quark binding, as in Sec.~\ref{SEC7}, as well as nucleon scattering and binding potentials \cite{Fried5:2012}.

Before proceeding with this phenomenological change, it is useful to describe the obstruction which will occur. As a typical and important part of a $4$-point fermionic function, one finds an exponential factor of
 \begin{eqnarray}\label{Eq:typic}
& & + \frac{i}{2} g \int{\mathrm{d}^{4}w_{} \, \int_{0}^{s}{\mathrm{d}s_{1} \, \int_{0}^{\bar{s}}{\mathrm{d}s_{2} \ u'_{\mu}(s_{1}) \, \bar{u}'_{\nu}(s_{2}) \,}}} \\ \nonumber & & \quad \times \, \Omega^{a}(s_{1}) \, \bar{\Omega}^{b}(s_{2}) \,  \left. (f \cdot \chi(w_{}))^{-1} \right|^{\mu\nu}_{ab} \\ \nonumber & & \quad \times \, \delta^{(4)}(w_{} - y_{1} + u(s_{1})) \, \delta^{(4)}(w_{} - y_{2} + \bar{u}(s_{2})).
\end{eqnarray}

\noindent This expression, corresponding to the interaction of particles 1 and 2, is obtained in the approximation of quenching and by neglecting quark's spins: We emphasize that the full non-approximate expression dsiplays exactly the same forms as the one under consideration,  which is why the point can be made using this simplified example.

In (\ref{Eq:typic}), $u_{\mu}$, $\Omega_{\mathrm{I}}^{a}$, and $s_{1}$ are variables associated with $\mathbf{G}_{c}^{\mathrm{I}}(x_{1}, y_{1}| A)$, whereas $\bar{u}_{\nu}$, $\bar{\Omega}_{\mathrm{I\!I}}^{b}$, and $s_{2}$ refer to $\mathbf{G}_{c}^{\mathrm{I\!I}}(x_{2}, y_{2}| A)$.  The last line of (\ref{Eq:typic}) may be written as
\begin{eqnarray}\label{Eq:39}
\delta^{(4)}(w - y_{1} + u(s_{1})) \, \delta^{(4)}(y_{1} - y_{2} + \bar{u}(s_{2}) - u(s_{1}))
\end{eqnarray}

\noindent and one sees that, as a consequence of EL, this interaction is peaked at $w_{0} = y_{1} - u(s_{1})$.  This means that all the other $w \neq w_{0}$ are irrelevant to the interaction, and, as discussed above, are removed along with their normalization factors, leaving dependence only upon a functional integration at one single point $w_0$.  However, use of the Image Measure Theorem~\cite{bctg:2012} can convert this functional integral into an ordinary, finite-dimensional integral, hereinafter denoted by $\int{\mathrm{d}^{n} \chi(w_{0})}$.

The point central to the argument of this Section has now being reached, as one evaluates the support of the second delta-function in (\ref{Eq:39}), which may be expressed as the product of delta-functions, in time, longitudinal and transverse coordinates,
\begin{eqnarray}\label{Eq:40}
& & \delta(y_{10} - y_{20} + \bar{u}_{0}(s_{2}) - u_{0}(s_{1}))  \\ \nonumber & & \quad \times \, \delta(y_{1\mathrm{L}} - y_{2\mathrm{L}} + \bar{u}_{\mathrm{L}}(s_{2}) - u_{\mathrm{L}}(s_{1})) \\ \nonumber & & \quad \times \, \delta^{(2)}(\vec{y}_{1\perp} - \vec{y}_{2\perp} + \vec{\bar{u}}_{\perp}(s_{2}) - \vec{u}_{\perp}(s_{1})).
\end{eqnarray}

\noindent In the CM of quark 1 and quark 2, one can choose the origin of each time coordinate as the time of their closest approach, and then the time difference $y_{10} - y_{20}$ is always zero.  If the $\mathrm{Q}_{1}$ and $\mathrm{Q}_{2}$ are scattering, then $y_{1\mathrm{L}} + y_{2\mathrm{L}} = 0$, since their longitudinal projections are in opposite directions; alternatively, if the $\mathrm{Q}_{1}$ and $\mathrm{Q}_{2}$ are bound together, then $y_{1\mathrm{L}} = y_{2\mathrm{L}}$, and their difference vanishes.  Either choice makes no difference at all to the following analysis, and so we adopt the simplest, second possibility, $y_{1\mathrm{L}} - y_{2\mathrm{L}} = 0$.
\par
But then, how should one interpret such a factor as
 \begin{eqnarray}\label{Eq:11'}
\delta( \bar{u}_0(s_{2}) - u_0(s_{1})) \, ?
\end{eqnarray}
At face value, at any given couple of values $(s_1, s_2)\in\  ]0, {s}]\times ]0, \bar{s}]$, and any pair of arbitrary functions $(u, {\bar{u}})$, each belonging to some infinite dimensional functional space, the probability of coincidence of $u(s_1)$ with ${\bar{u}}(s_2)$ is likely to be infinitesimally small if not zero.

Consider the time-coordinate delta-function, $\delta(\bar{u}_{0}(s_{2}) - u_{0}(s_{1}))$, which can have a zero argument whenever $\bar{u}_{0}(s_{2})$ and $u_{0}(s_{1})$ coincide.  Assume this happens at a set of points $s_{l}$, so that
\begin{eqnarray}
& & \delta(\bar{u}_{0}(s_{2}) - u_{0}(s_{1}))  \\ \nonumber &=& \sum_{\ell}{\delta(\bar{u}_{0}(s_{\ell}) - u_{0}(s_{1}) + (s_{2} - s_{\ell}) \cdot \bar{u}'_{0}(s_{\ell}) + \cdots)} \\ \nonumber &=& \sum_{\ell}{ \left. \frac{1}{|\bar{u}'_{0}(s_{\ell})|} \, \delta(s_{2} - s_{\ell}) \right|_{\bar{u}_{0}(s_{\ell}) = u_{0}(s_{1})}}.
\end{eqnarray}

\noindent In a similar way, the longitudinal delta-function may be evaluated as
\begin{eqnarray}
\sum_{m}{ \left. \frac{1}{|u'_{\mathrm{L}}(s_{m})|} \, \delta(s_{1} - s_{m}) \right|_{u_{\mathrm{L}}(s_{m}) = \bar{u}_{\mathrm{L}}(s_{2}) \rightarrow  \bar{u}_{\mathrm{L}}(s_{\ell})}},
\end{eqnarray}

\noindent and their product as
\begin{eqnarray}
\sum_{\ell , m}{ \frac{1}{|u'_{\mathrm{L}}(s_{m})|} \, \frac{1}{|\bar{u}'_{0}(s_{\ell})|} \, \delta(s_{1} - s_{m}) \, \delta(s_{2} - s_{\ell}) }
\end{eqnarray}

\noindent under the restrictions $u_{0}(s_{m}) = \bar{u}_{0}(s_{\ell})$ and $u_{\mathrm{L}}(s_{m}) = \bar{u}_{\mathrm{L}}(s_{\ell})$.  Now, $u_{0}$ and $\bar{u}_{0}$, and $u_{\mathrm{L}}$ and $\bar{u}_{\mathrm{L}}$ are continuous but otherwise completely arbitrary functions: The probability that the intersections of $u_{0}(s_{1})$ with $\bar{u}_{0}(s_{2})$ and of $u_{\mathrm{L}}(s_{1})$ with $\bar{u}_{\mathrm{L}}(s_{2})$ occur at exactly the same points is therefore arbitrarily small.  The only place where all four of these continuous functions have the same value is at $s_{1} = s_{2} = 0$, where, by definition of these functions, $u_{\mu}(0) = \bar{u}_{\mu}(0) = 0$, and hence, this pair of delta-functions collapse to the simple product,
\begin{eqnarray}\label{Eq:41'}
 \delta(\bar{u}_{0}(s_{2}) - u_{0}(s_{1}))\delta( \bar{u}_{3}(s_{2}) - u_{3}(s_{1}))=\frac{1}{ 2}\left(\frac{ \delta(s_{1})\delta(s_{2})}{ |u'_{3}(s_1)|  |\bar{u}'_{0}(s_2)|}+\frac{ \delta(s_{1})\delta(s_{2})}{|u'_{0}(s_1)|  |{{\bar{u}}}'_{3}(s_2)|}\right).
\end{eqnarray}

\noindent Note that in Wiener functional space \cite{Lapidus2000}, a proof of (\ref{Eq:41'}) can be given by the cogent form of Theorem \cite{hftg:2012,11tg:2011}. Note also that because of the factors appearing in the denominator, when the large quark's momenta ${P_1}_\mu$ and ${P_2}_\nu$ are substituted for the derivatives of the Fradkin's fields $u'_{\mu}(s_{1})$ and $\bar{u}'_{\nu}(s_{2})$, then the automatic suppression of the momenta dependence in (\ref{Eq:35}) becomes obvious since, up to a sign, they compensate for the very same factors in the first line of (\ref{Eq:typic}).

By definition one has
\begin{equation}\label{Fradkin}
u_{\mu}(s) = \int_{0}^{s}{\mathrm{d}s' \, u'_{\mu}(s')}, \quad u_{\mu}(0) = 0,
\end{equation}

\noindent then it is the remaining, transverse delta-function of (\ref{Eq:40}) which is now most relevant, the term
\begin{equation}\label{delta2b}
\delta^{(2)}(\vec{y}_{1\perp} - \vec{y}_{2\perp}) = \delta^{(2)}(\vec{b}),
\end{equation}

\noindent where $\vec{b}$ denotes the impact parameter, or transverse distance between the two scattering particles.  This $\delta^{(2)}(\vec{b})$ appears in the exponential of (\ref{Eq:34}), and the question arises as to what meaning it can be assigned.  Depending on its argument, a delta-function is either zero or infinite. In the first case this means that there is no interaction, while the second case means that at $\vec{b}=0$, one has an infinite phase factor, suggestive of hard disc scattering~\cite{Schiff}.

The relevant question is therefore why such a delta-function $\delta^{(2)}(\vec{b})$ appears at all.  The answer is that the assumption has earlier been made, in the conventional abelian way, that the quark and/or anti-quark may be treated as ordinary particles, whereas asymptotic quarks exist only in bound states: Their transverse coordinates cannot, in principle, be specified, and there is therefore no reason to retain the conventional (Abelian) practice in which such measurement is assumed possible. This is the interpretation that will be adopted here, taking the $\delta^{(2)}(\vec{b})$ outcome as a serious warning that some form of quark transverse fluctuation is necessary.

Why does this happens in QCD?  Because QCD possesses EL, which conventional abelian theories do not. The latter display sums over interconnected propagators, which provide a certain vagueness of position, whereas in the exact non-perturbative QCD, as described above, one finds the sharp determination of delta-functions corresponding to the EL property, and transverse imprecision must therefore be introduced separately, as a fundamental input to the theory.

In Ref.~\cite{Fried2009_QCD1}, it was suggested that this difficulty be treated in an \emph{ad hoc} phenomenological way, by replacing $\delta^{(2)}(\vec{b})$ by the smoothly varying, effective Gaussian
\begin{eqnarray*}
\varphi(\vec{b})=(2\pi)^{-2} \, \int{\mathrm{d}^{2}\vec{k} \, e^{i \vec{k} \cdot \vec{b} - \frac{\vec{k}^{2}}{4\mu^{2}}}},
\end{eqnarray*}

\noindent where $\mu$ is a mass parameter on the order of the Q-$\bar{\mathrm{Q}}$ bound state (which we shall call a "model pion"), although we were able to obtain the conclusions of that paper without specifying the precise form of $\varphi(\vec{b})$.  In this article, we face this question directly, by first developing a formalism in which transverse quark coordinates cannot be specified, and then showing how this formalism removes all such absurdities, such as that of the exponential factor of $\delta^{(2)}(\vec{b})$ above.  But it must be emphasized that our prescription is phenomenological, for there remains to be shown how such an approach could be derived from a more fundamental, operator-field version of QCD, in which transverse fluctuations would occur automatically, perhaps in relation to a possible non-commutative geometrical phase of non-perturbative QCD.

We would like to point out the scale change where the integral $\int{\mathrm{d}^{4}w \, \chi^{2}(w)}$ in the exponent of the Halpern representation (\ref{Eq:17}) is broken up into small cells of volume $(\delta_{ph})^4$,
\begin{equation*}
\int{\mathrm{d}^{4}w \, \chi^{2(w)} } \rightarrow  (\delta_{ph})^{4} \sum_{i}{\chi_i^2}, \quad \chi_{i} \equiv \chi(w_{i}).
\end{equation*}

\noindent Upon rescaling, $\chi_i\rightarrow (\delta_{ph})^4\chi'_i$, and re-expressing all interactions in terms of $\chi'$, there appears in (\ref{Eq:typic}) the factor $(\delta_{ph})^2\varphi(b)$.  In Ref.~\cite{Fried2009_QCD1} where transverse imprecision was treated in an {\it{ad hoc}} way, the size of $\delta_{ph}$ was taken to be $M^{-1}$, where $M$ corresponded to a very large energy associated with the eikonal limit. Here, we ask the more physical question of just how small that $\delta$ may be chosen in the light of an actual measurement, and we let Quantum Physics provide the answer: That contribution to the $\delta_{ph}$ corresponding to a time separation should be chosen as $1/E$, that corresponding to a (CM) longitudinal coordinate should be $1/p_{\mathrm{L}} \simeq 1/E$, while that corresponding to each of the transverse coordinates should be $1/\mu$; and hence the physical volume $(\delta_{ph})^{4}$ is proportional to $1/(\mu E)^2$. An alternate way of expressing this is that, starting from arbitrarily small separations in each coordinate, we average each $\chi_{i}$ variable over a physically meaningful distance, and call that average the $\chi_{i}$ contained in the volume $(\delta_{ph})^4$.  When a scale change $(\delta_{ph})^{2} \chi_{i} = \bar{\chi}_{i}$ is subsequently made in (\ref{Eq:51}), a factor of $(\delta_{ph})^{2}$ will appear multiplying $\varphi(b)$.

\section{\label{SEC5}A Phenomenological Expression of Transverse Imprecision}

Perhaps the simplest way of introducing transverse fluctuation is to average that part of the QCD Lagrangian dealing with the quark-gluon interaction, so that the transverse position of the color-charge current operator $\bar{\psi} \, \gamma_{\mu} \tau^{a} \, \psi(x)$ should be averaged over a small range by means of an initially unspecified distribution.  One can also demand the same imprecision for the vector current $\bar{\psi} \, \gamma_{\mu} \, \psi(x)$ and scalar density $\bar{\psi} \psi(x)$, but these extra requirements seem to complicate the presentation, to no real advantage, and will not be considered here.

We emphasize that we have here chosen perhaps the simplest phenomenological way of introducing a measure of such transverse imprecision/fluctuation; and that there are quite possibly other, more profound methods of obtaining the same objective.  In fact, we think it most probable that a lack of anti-commutation of quark operator fields at equal times, but at differing transverse coordinates could be the basic point requiring attention.  In the interests of simplicity and clarity, we ask the readers indulgence for postponing that particular investigation.

Instead of the conventional quark-gluon contribution to the Lagrangian density,
\begin{eqnarray}\label{Eq:42}
\mathcal{L}_{\mathrm{QG}} = - \bar{\psi} \, [m + \gamma_{\mu} \, (\partial_{\mu} - i g A_{\mu}^{a} \tau^{a})] \, \psi,
\end{eqnarray}

\noindent in which all field operators occur at the same space-time point, and for which gauge invariance under the standard QCD gauge transformations is obvious, we now adopt a local -- in time and longitudinal position -- but non-local in its transverse coordinates replacement,
\begin{eqnarray}\label{Eq:43}
\mathcal{L}'_{\mathrm{QG}} = ig \int{\mathrm{d}^{2} \vec{x}'_{\perp} \, \mathfrak{a}(\vec{x}_{\perp} - \vec{x}'_{\perp}) } \,  \bar{\psi}(x') \, \gamma_{\mu} A_{\mu}^{a}(x) \tau^{a} \, \psi(x'),
\end{eqnarray}

\noindent where the transverse imprecision function (TIF) $\mathfrak{a}(\vec{x}_{\perp} - {\vec{x}}'_{\perp})$ is a real, symmetric function of its arguments, of significant value only for distances on the order of the inverse of the pion mass, $x'_{\mu} = (x_{0}, \vec{x}'_{\perp}, x_{\mathrm{L}})$, and $A_{\mu}^{a}(x)$ is left untouched.  In this formulation, rigorous local gauge-invariance is suppressed for the underlying quark fields, whose quanta have unmeasurable transverse positions, but the hadrons all constructed from these quanta will nevertheless be proper singlets under SU(3).
\par
In fact, one may argue that required gauge invariance is maintained  when quark properties are averaged over distances below which no measurement of their transverse properties can be physically performed. It would be most attractive, if and when a better formalism is invented, if strict gauge invariance could be maintained for all values of transverse coordinates; but that is not a physical requirement, rather a mathematical nicety.

One notes that in the contribution of (\ref{Eq:43}) to its part of the Action operator, $\int{\mathrm{d}^{4}x \, \mathcal{L}_{\mathrm{QG}} }$, the $\vec{x}_{\perp}$ and $\vec{x}'_{\perp}$ coordinates can be interchanged, which yields an equivalent form in which every $A_{\mu}^{a}(x)$ of the original (\ref{Eq:42}) is replaced by $\int{\mathrm{d}^{2}\vec{x}'_{\perp} \, \mathfrak{a}(\vec{x}_{\perp} - \vec{x}'_{\perp}) \, A_{\mu}^{a}(x') }$. This interchange allows a very simple extraction of all such transverse imprecision, since both delta-functions of (\ref{Eq:39}) will now be replaced by
\begin{eqnarray}\label{Eq:44}
& & \int{\mathrm{d}^{2} {\vec{y}}_{1\perp}^{\, '} \, \mathfrak{a}(\vec{y}_{1\perp} - \vec{y}_{1\perp}^{\, '}) \, \int{\mathrm{d}^{2}\vec{y}^{\, '}_{2\perp} \, \mathfrak{a}(\vec{y}_{2\perp} - \vec{y}^{\, '}_{2\perp}) \, }}\\ \nonumber & & \quad \times  \delta^{(4)}(w - y'_{1} + u(s_{1})) \,  \delta^{(4)}(w - y'_{2} + \bar{u}(s_{2})).
\end{eqnarray}

\noindent One small complication of this procedure is that such 'primed' transverse coordinates will now appear in the arguments of $\chi$, \emph{e.g.}, $\vec{w}_{\perp} \rightarrow \vec{y}^{\, \prime}_{\perp}$, and the fixed position coordinate $\vec{w}_{\perp}$ must now itself be varied.  But the difference $|\vec{y}_{\perp} - \vec{y}^{\, \prime}_{\perp}|$ is effectively bounded by $1/\mu$, where $\mu$ appears in the definition of $\varphi(b)$ below, and it turns out that a negligible error is made when $\vec{w}_{\perp}$ is replaced by $\vec{y}_{\perp}$.  More details are given in Appendix~\ref{AppB}, where it is shown that this approximation is justified for the subsequent calculations of quark and nucleon bindings.

The first so modified delta-function of (\ref{Eq:44}) defines the argument $w_{1}$ of $\chi(w_{1})$, and we again observe that the final output of the Halpern FI will be an ordinary integral $\int{\mathrm{d}^{n}\chi}$, independent of the choice of $w_{1}$.  The second delta-function of (\ref{Eq:44}) now involves the $\mathfrak{a}$-dependence, generating in place of the $\delta^{(2)}(\vec{y}_{1\perp} - \vec{y}_{2\perp})$, the combination
\begin{eqnarray}\label{Eq:45}
& & \int{\mathrm{d}^{2}\vec{y}^{\, '}_{1\perp} \, \int{\mathrm{d}^{2}\vec{y}^{\, '}_{2\perp} \, \mathfrak{a}(\vec{y}_{1\perp} - \vec{y}^{\, '}_{1\perp}) \, \mathfrak{a}(\vec{y}_{2\perp} - \vec{y}^{\, '}_{2\perp}) }} \, \delta^{(2)}(\vec{y}^{\, '}_{1\perp} - \vec{y}^{\, '}_{2\perp}) \\ \nonumber &=& \int{\mathrm{d}^{2}\vec{y}^{\, '}_{\perp} \, \mathfrak{a}(\vec{y}_{1\perp} - \vec{y}^{\, '}_{\perp}) \, \mathfrak{a}(\vec{y}_{2\perp} - \vec{y}^{\, '}_{\perp})}.
\end{eqnarray}

\noindent Inserting 2-dimensional Fourier transforms of each
\begin{eqnarray}
\mathfrak{a}(\vec{y}_{\perp} - \vec{y}^{\, '}_{\perp}) = \int{\frac{\mathrm{d}^{2}\vec{k}}{(2\pi)^{2}} \, e^{i\vec{k} \cdot (\vec{y}_{\perp} - \vec{y}^{\, '}_{\perp})} \, \tilde{\mathfrak{a}}(\vec{k}_{\perp})},
\end{eqnarray}

\noindent the combination (\ref{Eq:45}) becomes
\begin{eqnarray}\label{Eq:46}
\int{\frac{\mathrm{d}^{2}\vec{k}}{(2\pi)^{2}} \, \tilde{\mathfrak{a}}(\vec{k}) \, \tilde{\mathfrak{a}}(-\vec{k}) \, e^{i\vec{k} \cdot (\vec{y}_{1\perp} - \vec{y}_{2\perp})} }.
\end{eqnarray}

\noindent From its definition, $\mathfrak{a}$ is real, and hence (\ref{Eq:46}) becomes
\begin{eqnarray}\label{Eq:47}
\int{\frac{\mathrm{d}^{2}\vec{k}}{(2\pi)^{2}} \, \, e^{i\vec{k} \cdot \vec{b}} \, \left|\tilde{\mathfrak{a}}(\vec{k})\right|^{2} } \equiv \varphi(\vec{b}),
\end{eqnarray}

\noindent which provides the definition of $\varphi(\vec{b})$.  Note that while no restriction has been placed on the form of $\mathfrak{a}$ other than that it is real and symmetric, $\varphi$ turns out to be independent of the direction of $\vec{b}$, that is, $\varphi(\vec{b})=\varphi(b)$.  In this way, the improper $\delta^{(2)}(b)$ is automatically replaced by $\varphi(b)$.

\section{\label{SEC6}Bundle Diagrams}

\begin{figure}
\includegraphics[height=35mm]{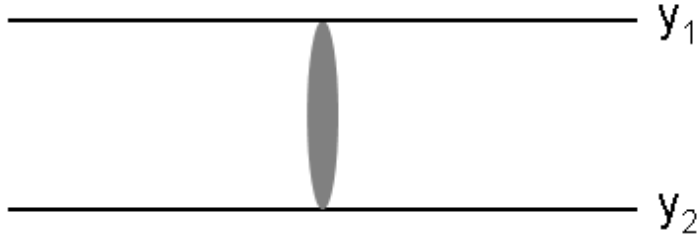}%
\caption{\label{Fig:1}A gluon bundle exchanged between quarks $\mathrm{I}$ and $\mathrm{I\!I}$}
\end{figure}

		In the above example of quark and/or antiquark scattering, where the infinite number of exchanged gluons appears to originate and end at a single space-time point on a quark/antiquark line, {\it{modulo}} transverse imprecision, it may be helpful to introduce the concept of an exchanged `gluon bundle', as in Figure~{\ref{Fig:1}}.  Because of the four-dimensional delta function $\delta^{(4)}(y'_{1} - y'_{2} - u(s_{1}) + \bar{u}(s_{2}))$, arising from the product of the pair of delta functions of (\ref{Eq:44}), and of the subsequent analysis which produces (\ref{Eq:45}), the transverse separation $\vec{b} = \vec{y}_{1\perp} - \vec{y}_{2\perp}$ satisfies the probability distribution (\ref{Eq:47}).  The argument $w$ in $(f \cdot \chi(w))^{-1}$ is given by $w = w_{1} = y'_{1} - u(s_{1}) = y'_{1} \rightarrow y_{1}$, in virtue of (\ref{Eq:41'}), and, as explained in Sec.~\ref{SEC3}, the Halpern functional integral reduces to a set of ordinary integrals \cite{bctg:2012} that are represented by the Bundle Diagram of Figure~{\ref{Fig:1}}.

Here, therefore, it is understood that time and longitudinal coordinates of the end-points of the bundle are the same, whereas their transverse coordinates, measured vertically in the figure, are separated.  Bundle diagrams are not Feynman diagrams, but offer perhaps a more efficient way of representing the sum over all of the Feynman graphs corresponding to such multiple gluon exchange.

\begin{figure}
\includegraphics[keepaspectratio,height=35mm]{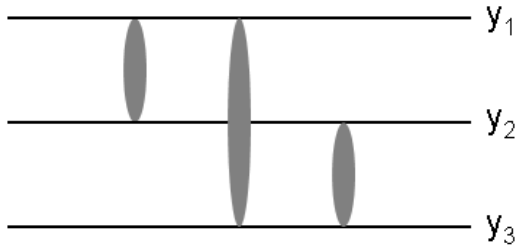}%
\caption{\label{Fig:2}Gluon bundles exchanged among three quarks.}
\end{figure}

A slightly more complicated expression describes gluon bundles exchanged between any two of three quarks, as in Figure~{\ref{Fig:2}}, where, because of EL, the $w$-coordinates of each of the $(f \cdot \chi)^{-1}$ entering into the appropriate Halpern functional integral are the same, even though the transverse coordinates of the three quarks can be quite different.

\begin{figure}
\includegraphics[keepaspectratio,height=35mm]{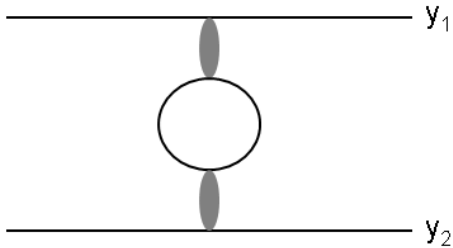}%
\caption{\label{Fig:3}Gluon bundles joining two quarks via a closed quark loop.}
\end{figure}

In contrast, were a closed quark loop, corresponding to a simple relaxation of the quenched approximation, to appear between a pair of quarks, joined to each external quark line by the exchange of a gluon bundle, as in Figure~{\ref{Fig:3}}, there will now be two distinct sets of ordinary Halpern integrals to be evaluated.

As will be seen elsewhere, the effective diagram of Figure~{\ref{Fig:3}} will provide us with the essential features of the Nucleon-Nucleon potential for separation lengths beyond 2 fm~\cite{Fried5:2012}.  Interestingly also, a comparison of bundle diagrams such as those depicted in Figures~\ref{Fig:1}, \ref{Fig:2} and \ref{Fig:3} will be shown to provide an interesting qualitative understanding of the difference between QCD and its pure Yang-Mills truncation.

%
%
%
\section{\label{SEC7}Quark Binding Potential}

This Section should be viewed as a continuation of that of Ref.~\cite{Fried2009_QCD1}, in which a new formalism for analytic QCD calculations was suggested, and which can serve to identify an effective potential as the result of the exchange of multiple virtual gluons between $\mathrm{Q}$'s and/or $\bar{\mathrm{Q}}$'s.  The word 'effective' here represents the type of potential one would expect to find were the situation that of a scattering of conventional Abelian particles, moving at large relative velocities in their CM, as appropriate in an eikonal context.  The word 'Abelian' is relevant in this QCD situation, for the analysis of Ref.~\cite{Fried2009_QCD1} suggests that for large impact parameters the scattering of such non-Abelian $\mathrm{Q}$'s and/or $\bar{\mathrm{Q}}$'s is effectively coherent; only as the impact parameter decreases do progressively stronger color fluctuations appear and decrease the effective coupling, which then becomes the statement of non-perturbative asymptotic freedom in the variable conjugate to the impact parameter.  For such scattering, the eikonal model is, by far, the simplest method of extracting the relevant Physics.

This and the next Section represent two, simple, phenomenological extensions of Sec.~\ref{SEC5}.  However, they are supported by a more rigorous analysis~\cite{hftg:2012} where, using quenched and eikonal approximations, the exact closed forms of such QCD fermionic amplitudes are obtained.

We shall be interested in binding, or "restoring" potentials, effective potentials which prevent bound particles from remaining at relatively large separations.  Since large impact parameters are relevant to large, three-dimensional distances, what is needed is, firstly, a statement of the relevant formulae of Ref.~\cite{Fried2009_QCD1} at large impact parameter; and, secondly, the corresponding eikonal function found in that limit.  It is then a relatively simple matter to determine an effective potential which would produce the same eikonal.

A two-body eikonal scattering amplitude will have the form~\cite{HMF2}
\begin{eqnarray}\label{Eq:50}
\mathcal{T}(s,t) = \frac{is}{2m^{2}} \int{\mathrm{d}^{2}\vec{b} \, e^{i \vec{q} \cdot \vec{b}} \left[ 1 - e^{i\mathbf{X}(s,\vec{b})}\right]},
\end{eqnarray}

\noindent where $s$ and $t$ denote the standard Mandelstam variables, $s = -(p_{1} + p_{2})^{2}$, $t = -(p_{1} - p'_{1})^{2} = \vec{q}^{\, 2}$ in the CM of $\mathrm{Q}\bar{\mathrm{Q}}$, and where $\mathbf{X}(s,b)$ is the eikonal function.  Following the arguments of \cite{Fried2009_QCD1}, we give here the simplest derivation of results whose qualitative form describes that of a more accurate analysis \cite{hftg:2012}.  One reaches
\begin{eqnarray}\label{Eq:51}
e^{i\mathbf{X}} = \mathcal{N} \int{\mathrm{d}^{n} \bar{\chi}_{30} \, \left[ \det(g f \cdot \bar{\chi})^{-1}\right]^{\frac{1}{2}} \cdot e^{i \frac{\bar{\chi}^{2}}{4}} \cdot \exp{\left[ i g \varphi(b) \, \Omega^{a}_{\mathrm{I}} \cdot \left. (f \cdot \bar{\chi})^{-1}\right|^{ab}_{30} \cdot \Omega^{a}_{\mathrm{I\!I}} \right]  }},
\end{eqnarray}

\noindent and the normalization is such that for $g\varphi(b) \rightarrow 0$, $\exp{[i \mathbf{X}]} \rightarrow 1$. Here, $n=N_c^2-1$, that is $n=8$ at $N_c=3$.  All the integrals over
\begin{eqnarray}\label{Eq:52}
\int{\mathrm{d}^{n} \alpha_{\mathrm{I}}} \, \int{\mathrm{d}^{n} \alpha_{\mathrm{I\!I}}} \, \int{\mathrm{d}^{n} \Omega_{\mathrm{I}}} \, \int{\mathrm{d}^{n} \Omega_{\mathrm{I\!I}}}
\end{eqnarray}

\noindent are properly normalized, and they connect the $\Omega^{a}_{\mathrm{I}}$, $\Omega^{b}_{\mathrm{I\!I}}$ dependence of (\ref{Eq:51}) with the $\lambda^{a}_{\mathrm{I}}$, $\lambda^{b}_{\mathrm{I\!I}}$ Gell-Mann matrices placed between initial and final state vectors, which define the type of amplitude desired, $\mathrm{Q}$-$\mathrm{Q}$ scattering or $\mathrm{Q}$-$\bar{\mathrm{Q}}$ scattering, \emph{etc.}

As in \cite{Fried2009_QCD1}, the 8-dimensional integral over $\int{\mathrm{d}^{n} \bar{\chi}_{03}}$ is arranged into the form of a radial integration over the magnitude of $\bar{\chi}_{03}$, now called $R$, and normalized angular integrations.  We suppress the latter, along with all normalized $ \int{\mathrm{d}^{n} \Omega_{\mathrm{I}}} \, \int{\mathrm{d}^{n} \Omega_{\mathrm{I\!I}}}$ dependence, and rewrite (\ref{Eq:51}) in the form
\begin{eqnarray}\label{Eq:53}
e^{i\mathbf{X}(s,b)} = \mathcal{N}' \int_{0}^{\infty}{\mathrm{d}R \, R^{3} \, e^{i \frac{R^{2}}{4} + i  \frac{\langle g\rangle\varphi(b)}{R} } },
\end{eqnarray}

\noindent where the normalization of the $R$-integral is given by ${\mathcal{N}'}^{-1} = \int_{0}^{\infty}{\mathrm{d}R \, R^{3} \, e^{i \frac{R^{2}}{4} }}$, and where $\langle g \rangle$ is a shorthand denoting the effective coupling after all angular and color integrations and matrix elements have been calculated.  In this calculation, only the linear dependence of $g \varphi$ is needed, while the non-vanishing matrix elements between singlet color states will define $\mathrm{Q}$-$\bar{\mathrm{Q}}$ scattering. The integer power of $3$ in (\ref{Eq:53}), of the term $R^3=R^{7-4}$, arises from the $\frac{1}{2}Tr\ln\hat{K}$ factor of (\ref{Eq:34}), contributing $R^{-4}$ to (\ref{Eq:51}).  As noted above, we wish to compare this eikonal function at fairly large impact parameter with that corresponding to the scattering of two, interacting particles, and then infer what form of effective potential corresponds to that eikonal. We therefore assume the usual formula~\cite{HMF2} relating an eikonal to a specified potential,
\begin{eqnarray}\label{Eq:54}
\mathbf{X}(s,b) = - \int_{-\infty}^{\infty}{\mathrm{d}z \, V(\vec{b}+z \hat{\mathrm{p}}_{\mathrm{L}})} = \gamma(s) \, \mathbf{X}(b),
\end{eqnarray}

\noindent where $\gamma(s)$ depends upon the CM energy of the scattering particles (and the nature of their interaction), and $\hat{\mathrm{p}}_{\mathrm{L}}$ is a unit vector in the direction of longitudinal motion.  Equation (\ref{Eq:54}) is true for non-relativistic and relativistic scattering, in potential theory and in field theory. For large $b$, our eikonal turns out to be weakly dependent on $s$, and the $b$-independent contributions do not contribute to $V(r)$.

To obtain $V(r)$ from a given  $\mathbf{X}(b)$ is then simply a matter of reversing the usual calculation: One computes the Fourier transform $\tilde{\mathbf{X}}(k_{\perp})$ of  $\mathbf{X}(b)$, and then extends $k_{\perp}$ to $|\vec{k}| = [ \vec{k}_{\perp}^{2} + k_{\mathrm{L}}^{2} ]^{\frac{1}{2}}$.  The three-dimensional Fourier transform of $\tilde{\mathbf{X}}(k)$ is then proportional to $V(r)$.  Following Sec.V, we assume that $\mathbf{X}(b)$, which depends on the form of $\varphi(b)$, is a function of ${\vec{b}}^{2}$; and although we begin by asking for the form of $\varphi(b)$ for large $b$, it can be shown that the small-$b$ contributions to the Fourier transforms are unimportant for the large-$r$ behavior of $V(r)$.

One further point requires consideration, for the above remarks are valid when the process involves only the scattering of two particles.  But when initial energies are high enough such that inelastic particle production occurs, then the potential corresponding to such production must have a negative imaginary component (so that the corresponding S-matrix, $ \exp{[-i\mathcal{H}t]}$, will have a decaying time dependence), thereby conserving probability as a diminution of the final, two-particle state when production becomes possible. This is also true when the two initial particles have the possibility of binding, and forming a state which was not initially present; the probability of the final scattering state must diminish.  In other words, the eikonal function one expects to find may always be characterized by a complex potential of form $V_{\mathrm{R}} - i V_{\mathrm{I}}$, where in the present case, $V_{\mathrm{I}}$ is that potential which can bind the $\mathrm{Q}$ and $\bar{\mathrm{Q}}$ into a pion:  In a scattering calculation, that potential must therefore appear with a multiplicative factor of $- i$ (symbolically, $i\mathbf{X} \rightarrow i(V_{\mathrm{R}} - i V_{\mathrm{I}})$, and for this calculation of binding, $i\mathbf{X} \rightarrow V_{\mathrm{I}}=V_{\mathrm{B}}$).

Returning to (\ref{Eq:50}), it is assumed that $\vec{q} \neq 0$, so that the "1" term of the integrand contributes a $\delta^{(2)}(q)$ that can be discarded.  This amounts to a physical assumption: $\mathrm{Q}$ and $\bar{\mathrm{Q}}$ are not bound rigidly, with an unchanging impact parameter; rather, there is a permanent, if small, "in-and-out" transverse motion, which represents the bound state, and hence $\vec{q} \neq 0$.  Then,
\begin{eqnarray}\label{Eq:55}
\mathcal{T}(s,t) \sim  \int{\mathrm{d}^{2}\vec{b} \  e^{i \vec{q} \cdot \vec{b}} \ e^{i\mathbf{X}(b)} },
\end{eqnarray}

\noindent where $\exp{[i \mathrm{X}(b)]}$ is given by (\ref{Eq:53}).  In \cite{Fried2009_QCD1} it was assumed that $\varphi(b)$ could be a smooth function which vanishes for large $b$, such as $\sim \exp{[-(\mu b)^{2}]}$, but it turns out that an appropriate choice is
\begin{equation}\label{Eq:levy}
\varphi(b) = \varphi(0) \, e^{- (\mu b)^{2+\xi} },
\end{equation}

\noindent where $\xi$ is real, positive and small.  For any such choice of $\xi$, $\varphi(b)$ becomes small as $\mu b \gg 1$.

For large $b$ it is then sensible to expand the $\exp{[i \langle g \rangle \varphi(b) / R]}$ term of (\ref{Eq:53}), retaining only the linear $\langle g \rangle \varphi$ dependence.  This gives, in place of (\ref{Eq:53}),
\begin{eqnarray}\label{Eq:57}
e^{i\mathbf{X}(b)} &\simeq& \mathcal{N}' \int_{0}^{\infty}{\mathrm{d}R \, R^{3} \, e^{i \frac{R^{2}}{4}} \left[1 + i \langle g\rangle \frac{\varphi(b)}{R} + \cdots \right] } \\ \nonumber &=& 1 + i \kappa \mathcal{N}' \langle g\rangle \varphi(b) + \cdots,
\end{eqnarray}

\noindent where
\begin{eqnarray}\label{Eq:58}
\kappa \mathcal{N}' = \mathcal{N}' \int_{0}^{\infty}{\mathrm{d}R \, R^{3} \, e^{i \frac{R^{2}}{4}} } = 2 \sqrt{\pi} (-i)^{\frac{3}{2}},
\end{eqnarray}

\noindent and $\mathcal{N}' = -\frac{1}{8}$.  Remembering that both sides of (\ref{Eq:57}) are to be integrated over $\int{\mathrm{d}^{2}\vec{b} \, e^{i \vec{q} \cdot \vec{b}} }$, for $\vec{q} \neq 0$ we can proceed to the identification
\begin{equation}\label{Eq:59}
e^{i\mathbf{X}(b)} = i \kappa \mathcal{N}' {\langle g \rangle} \varphi(b) ,
\end{equation}

\noindent or
\begin{equation}\label{Eq:60}
i\mathbf{X}(b) = \ln{\left[ \varphi(b)\right]} + \cdots,
\end{equation}

\noindent where $\ln{\left[ \varphi \right]}$ is large and $\varphi$ is (effectively) small, and where the dots will be commented shortly.

Choosing $\varphi(b) = \varphi(0) \, e^{- (\mu b)^{2+\xi}}$, one obtains
\begin{equation}\label{Eq:61}
i\mathbf{X}(b) = - (\mu b)^{2+\xi} + \cdots .
\end{equation}

\noindent It is then convenient to use the integrals~\cite{GR}
\begin{equation}\label{Eq:62}
\int_{0}^{\infty}{\mathrm{d}x \, x^{\mu} \, J_{0}(ax)} = 2^{\mu} \, a^{-1-\mu} \, {\Gamma(\frac{1}{2} + \frac{\mu}{2})} / {\Gamma(\frac{1}{2} - \frac{\mu}{2})}, \quad \mu < \frac{1}{2},
\end{equation}

\noindent and
\begin{equation}\label{Eq:63}
\int_{0}^{\infty}{\mathrm{d}x \, x^{\mu - 1} \, \sin(x)} = \Gamma(\mu) \, \sin(\frac{\mu \pi}{2}), \quad |\Re{(\mu)}| < 1,
\end{equation}

\noindent in which, except for obvious poles, the Gamma functions are analytic in $\mu$, and can be continued to the needed values.  With
\begin{eqnarray}\label{Eq:64}
i\tilde{\mathbf{X}}(k_{\perp}) &=& - \int{\mathrm{d}^{2}b \, (\mu b)^{2+\xi} \, e^{i \vec{k}_{\perp} \cdot \vec{b}} } \\ \nonumber &=& - (2 \pi)\int_{0}^{\infty}{\mathrm{d}b \, J_{0}(k_{\perp} b) \, \mu^{2+\xi} \, b^{3+\xi} },
\end{eqnarray}

\noindent and the use of the doubling formula for Gamma functions, working everything through, one finds
\begin{equation}\label{Eq:65}
V_{\mathrm{B}}(r) = -\frac{2^{3+\xi}}{\pi} \, \mu^{2+\xi} \, r^{1+\xi} \, \frac{\Gamma(2 + \frac{\xi}{2})}{\Gamma(-1 - \frac{\xi}{2})} \, \Gamma(-2-\xi) \, \sin(\frac{\pi  \xi}{2}),
\end{equation}

\noindent and for small enough $\xi$, the confining potential
\begin{equation}\label{Eq:66}
V_{\mathrm{B}}(r) \simeq \xi \, \mu \, (\mu r)^{1+\xi},
\end{equation}

\noindent which can be compared to the results of several machine groups~\cite{Ref-Machine}. It is remarkable that the choice (\ref{Eq:levy}) can be viewed as part of a Levy-flight probability distribution \cite{levy} that generalizes Gaussian ones in consistency with the famous {\it{ Central Limit Theorem}} of statistical physics. It is interesting to note that at a physical level, such a distribution addresses the issue of transverse quark motions, describing them in a way that can seem quite relevant to a confined context.  Likewise, (\ref{Eq:levy}) could suggest a most fascinating connection to a non-commutative geometrical aspect of non-perturbative QCD, and this should be examined elsewhere.

There remains the question of what values should be assigned to $\xi$, and to the unknown parameter $\mu / m_{\mathrm{Q}}$. These issues will be dealt with in the next Section.  Incidentally, one may observe that in (\ref{Eq:60}), additive constants to $\ln{[\varphi(b)]}$ bring no contribution to $V_{\mathrm{B}}(r)$.

To calculate the corresponding effective restoring potential when one of the three quarks contributing to the $\mathrm{QQQ}$ bound state is separated from the other two, or, more simply, when all three quarks are forced to separate from each other, one may refer to Equation (22) of \cite{Fried1983}, giving the eikonal amplitude corresponding to Coulomb three-particle scattering.  Here, the specifically Coulomb parts of this amplitude may be replaced by the single Halpern integral which connects the three quarks to each other, as it connected the $\mathrm{Q}$ and $\bar{\mathrm{Q}}$ to each other in the calculation above.  If particle 2 of this formula enters the scattering with zero transverse momentum in the rest frame of particles 1 and 3, the amplitude simplifies to
\begin{eqnarray}\label{Eq:67}
\mathcal{T}_{3}^{\mathrm{eik}} \sim \int{\mathrm{d}^{2}\vec{b}_{12} \, \int{\mathrm{d}^{2}\vec{b}_{32} \, e^{i \vec{q}_{3} \cdot \vec{b}_{32} + i \vec{q}_{1} \cdot \vec{b}_{12} } \left[ \Phi(\vec{b}_{12}, \vec{b}_{32}, \vec{b}_{13}) + \Psi_{3} \right] }},
\end{eqnarray}

\noindent where $\vec{b}_{13} = \vec{b}_{12} - \vec{b}_{32}$, irrelevant kinematic factors multiplying these integrals have been suppressed, and $\Psi_{3}$ denotes the combination
\begin{equation}\label{Eq:68}
-1 + \left[ 1 - \Phi(\vec{b}_{12}, \infty, \infty) \right] + \left[ 1 - \Phi(\infty, \vec{b}_{32}, \infty) \right] + \left[ 1 - \Phi(\infty, \infty, \vec{b}_{13}) \right],
\end{equation}

\noindent so that $\mathcal{T}_{3}^{\mathrm{eik}}$ is a completely "connected" amplitude.

Writing $\Phi$ as $\exp{[i \mathbf{X}(b_{12}, b_{32})]}$, its defining integral may be rewritten as
\begin{equation}\label{Eq:69}
\mathcal{N}' \int_{0}^{\infty}{\mathrm{d}R \, R^{3} \, \exp{\left[ i \frac{R^{2}}{4} + i \left[ {\langle g \rangle}_{12} \varphi(b_{12}) + {\langle g \rangle}_{32} \varphi(b_{32}) + {\langle g \rangle}_{13} \varphi(b_{13}) \right]\right]} },
\end{equation}

\noindent and one may ask for the form (\ref{Eq:69}) takes when one or more of the $b_{ij}$ becomes large.

When any one of the $b_{ij}$ becomes large, $\varphi(b_{ij})$ becomes small, and its exponential term may be expanded.  The simplest situation is when they all become large, so that an expansion of (\ref{Eq:69}) is relevant.  The first non-zero and leading term in such an expansion which contains the $\varphi$ dependence, and hence the $b_{ij}$-dependence, of all three quarks contributing to the singlet $\mathrm{QQQ}$ state is that term for which the expansion yields
\begin{equation}\label{Eq:70}
1 + \kappa \mathcal{N}' \left[ {\langle g \rangle}_{12} \varphi(b_{12}) + {\langle g \rangle}_{32} \varphi(b_{32}) + {\langle g \rangle}_{13} \varphi(b_{13}) \right] + \cdots,
\end{equation}

\noindent so that, suppressing all normalized angular and color integrations, for $\vec{q}_{1} \neq 0$, $\vec{q}_{3} \neq 0$,
\begin{equation}\label{Eq:71}
e^{i \mathbf{X}(b_{12}, b_{32}) } \simeq \kappa \mathcal{N}' \left[ {\langle g \rangle}_{12} \varphi(b_{12}) + {\langle g \rangle}_{32} \varphi(b_{32}) + {\langle g \rangle}_{13} \varphi(b_{13}) \right] + \cdots.
\end{equation}

\noindent If particles 1 and 3 are now additionally separated, keeping the distances $b_{12}$ and $b_{32}$ essentially fixed, then
\begin{eqnarray}\label{Eq:72}
e^{i \mathbf{X}(b_{12}, b_{32}) } &\simeq& \kappa \mathcal{N}' {\langle g \rangle}_{13} \varphi(b_{13}) + \cdots, \\ \nonumber i \mathbf{X}(b_{12}, b_{32}) &\simeq& \ln{\left[\varphi(b_{13})\right]} + \cdots,
\end{eqnarray}

\noindent and we are effectively back in the "pion" situation, where the large-impact parameter eikonal of $\mathrm{Q}$-$\bar{\mathrm{Q}}$ scattering was calculated. Clearly, (\ref{Eq:72}) suggests that when any two of the three quarks which contribute to the singlet "nucleon" are well separated, there results a restoring potential $V(r_{ij})$ of the same form as that of (\ref{Eq:66}).  There are, of course, corrections to this confining potential, those which are obvious from the approximations used above, as well as those corresponding to spin and angular momentum, which have been completely neglected.

\section{\label{SEC8}Estimation of the "Model Pion" Mass}

This Section contains an estimate of the effects of the above binding potential for the simplest case of the "pion", a ground state of the $\mathrm{Q}$-$\bar{\mathrm{Q}}$ system calculated using the simplest minimization technique~\cite{Quantics JMLL}, as well as a second minimization with respect to a ratio of terms introduced in this analysis.
\par
In this simplest, non-relativistic estimation, the Hamiltonian of this system is given by
\begin{equation}\label{Eq:73}
\mathcal{H} = 2m + \frac{1}{m} p^{2} + V(r),
\end{equation}

\noindent where $1/m$ denotes the inverse of the reduced mass of this equal-mass quark system.  The ground-state energy is then given by the replacement of $p$ by $1/r$, and the subsequent minimization of the eigenvalue of this Hamiltonian with respect to $r$.  Assuming $\xi \ll 1$, one obtains
\begin{equation}\label{Eq:74}
\mu r = \left( \frac{2}{\xi} \, \frac{\mu}{m} \right)^{1/3}
\end{equation}

\noindent and
\begin{equation}\label{Eq:75}
E_{0} = 2m + \frac{3}{2} \xi \mu \left( \frac{2}{\xi} \, \frac{\mu}{m} \right)^{1/3}.
\end{equation}

\noindent Here, $\mu$ and $\xi$ are the two parameters required by this analysis in order to have a non-zero binding potential; and while their existence has been introduced as a necessary assumption for the interacting $Q$-$\bar{Q}$ system, there is another unknown quantity, the quark mass $m$, which must play an important role.  If we take the position that $\mu$ and $\xi$ will appear as a result of a more fundamental QFT (in which the transverse momenta or position coordinates of the quanta of the QCD fields can never be measured with precision), then we are free to consider which values of $m$, or of the ratio $\mu / m$, can lead to the lowest value of $E$.  In so doing, one is asking if there might exist a dynamical reason for the relatively small value of the pion mass, in this approximation to the actual pion; that is, whether approximate chiral symmetry has a dynamical basis.

Following this approach and using the small-$\xi$ limit of (\ref{Eq:75}),
\begin{equation}\label{Eq:76}
E_{0} = \mu \left[ 2 \left(\frac{\mu}{m}\right)^{-1} + 3 \left(\frac{\xi}{2}\right)^{2/3} \, \left( \frac{\mu}{m} \right)^{1/3} \right].
\end{equation}

\noindent Then, upon variation with respect to $x = \mu / m$, the function $E_{0}(x)$ will have an extremum at $x_{0} = 2^{3/4} \left( \frac{2}{\xi} \right)^{1/2}$, and since
\begin{equation}\label{Eq:77}
\left. \frac{\partial^{2} E_{0}(x)}{\partial x^{2}} \right|_{x_{0}} = \mu \, 2^{-5/4} \, \left(\frac{\xi}{2}\right)^{3/2} \, \left( 2 - \frac{2}{3} \right) > 0,
\end{equation}

\noindent  that extremum is a minimum.  Substituting the value of $\mu / m = 2^{3/4} \, \left(\frac{2}{\xi}\right)^{1/2}$ into (\ref{Eq:76}) yields
\begin{equation}\label{Eq:78}
E_{0} = \mu \, \xi^{1/2} \, 2^{-1/4} \, [1 + 3],
\end{equation}

\noindent from which one infers that there is three times as much energy in the gluon field and $\mathrm{Q}$-$\bar{\mathrm{Q}}$-kinetic energies as in the quark rest masses.  Intuitively, one expects that $E \sim m_{\pi} \sim \mu$, which suggests that $\xi \sim \sqrt{2}/16$.  And finally, one then has an estimate of the quark mass, in terms of $\xi$ and $\mu$.
%
%

Of course, this double-minimization calculation of $E_{0}$ cannot be exactly identified with the precise pion mass, because of the approximations made above, as well as the omission of more complicated singlet terms, such as the contributions coming from $\mathrm{Q} \mathrm{Q} \bar{\mathrm{Q}} \bar{\mathrm{Q}}$ terms.  But the Physics seems to be reasonably correct.
Note also that we are dealing with "textbook" QCD, containing but one type of quark and its complement of massless, SU(3) gluons; flavors and electro-weak corrections can be added, as desired.

A similar minimization analysis can be made for the nucleon ground state, but there then are other degrees of freedom which should be included, and treated properly by a serious attempt at a solution of such a three-body, relativistic problem. This we must leave to others whose numerical abilities far exceed our own.

%
%
%
\section{\label{SEC9}Summary and Speculation}

The previous Sections have presented the formulation of a new approach to analytic, non-perturbative, gauge-invariant QCD, from the first step of demonstrating how gauge invariance of the Schwinger/Symanzik generating functional can be assured; to the explicit demonstration of how that invariance is achieved by gauge-independence of the non-perturbative sums over all gluons exchanged between scattering quarks and/or antiquarks; to the expression of a new property of non-perturbative QCD called Effective Locality, and the necessity of introducing 'transverse imprecision' in the basic Lagrangian of QCD; and ending with an explicit determination of the quark-binding potential, and its use in estimating a model pion.

All of this has been accomplished in the context of a 'textbook' QCD example, involving but one type of massive quark and its complement of massless SU(3) gluons. For ease of presentation, a most convenient eikonal model has been used for high-energy processes, although all steps could have been carried out without approximation by the use of an exact Fradkin represenation for $\mathbf{G}_{\mathrm{c}}[A]$.  Flavors and electro-weak interactions can be included when desired.  The method of assuring gauge invariance, although simple, had been overlooked for decades; while the EL property, along with the necessity of transverse fluctuations, are,  to the best of our knowledge, new requirements inherent to non-perturbative QCD.

As an indication of the usefulness of this approach, a subsequent paper now under preparation will derive from 'realistic' QCD, a quite good representation of a nucleon-nucleon binding potential, as appropriate to the deuteron. This, to our knowledge at least, is the first analytic QCD derivation relevant to the entire subject of Nuclear Physics, while it also suggests obvious glueball candidates. That calculation requires the use of a closed-quark-loop, as in Fig.~\ref{Fig:3}; and the subject of further 'radiative corrections', and indeed, of non-perturbative QCD renormalization, makes clear that the adjective 'non-perturbative' used in this paper refers only to the complete summation of all possible gluon exchanges between quarks and/or anti-quarks.  The questions then arises: Is it possible to extend this formalism to include sums over all closed quark loops?  What is the effect of gluon bundle exchange within and between closed quark loops?  Is it possible to achieve a totally non-perturbative statement of any given QCD amplitude?  We hope to answer these and other, related questions in subsequent publications.

\begin{acknowledgments}
This publication was made possible through the support of a Grant from the John Templeton Foundation. The opinions expressed in this publication are those of the authors and do not necessarily reflect the views of the John Templeton Foundation.  We especially wish to thank Mario Gattobigio for his many kind and informative conversations relevant to the Nuclear Physics aspects of our work.  It is also a pleasure to thank Mark Restollan, of the American University of Paris, for his kind assistance in arranging sites for our collaborative research when in Paris.
\end{acknowledgments}

\appendix

\section{\label{AppA}Fradkin's Representations for $\mathbf{G}_{\mathrm{c}}[A]$ and $\mathbf{L}[A]$}

The exact functional representations of these two functionals of $A(x)$ are perhaps the most useful tools in all of QFT, for they allow that $A$-dependence of these functionals to be extracted from inside ordered exponentials; and because they, themselves, are Gaussian in their dependence upon $A(x)$, they permit the functional operations of the Schwinger/Symanzik generating functional (Gaussian functional integration, or functional linkage operation) to be performed exactly.  This corresponds to an explicit sum over all Feynman graphs relevant to the process under consideration, with the results expressed in terms of functional integrals over the Fradkin variables; and in the present QCD case, because of EL, those non-perturbative results can be extracted and related to physical measurements.

The causal quark Green's function (which is essentially the most customary Feynman one) can be written as~\cite{Fradkin1966,HMF2}
\begin{equation}\label{Eq:A1}
\mathbf{G}_{c}[A] = [ m  + i \gamma \cdot \Pi][m  + (\gamma \cdot \Pi)^{2}]^{-1} = [ m  + i \gamma \cdot \Pi] \cdot i \int_{0}^{\infty}{ds \, e^{-ism^{2}} \, e^{is(\gamma \cdot \Pi)^{2}}  },
\end{equation}

\noindent where $\Pi = i [\partial_{\mu} - i g A_{\mu}^{a} \tau^{a}]$ and $(\gamma \cdot \Pi)^{2} = \Pi^{2} + i g \sigma_{\mu \nu} \, \mathbf{F}_{\mu \nu}^{a} \tau^{a}$ with $\sigma_{\mu \nu} = \frac{1}{4} [\gamma_{\mu}, \gamma_{\nu}]$.  Following Fradkin's method \cite{Fradkin1966, HMF2} and replacing $\Pi_{\mu}$ with $i \frac{\delta}{\delta v_{\mu}}$, one obtains
\begin{eqnarray}\label{Eq:A2}
& & \mathbf{G}_{\mathrm{c}}(x,y|A) \\ \nonumber &=& i \int_{0}^{\infty}{ds \ e^{-ism^{2}} \cdot e^{i \int_{0}^{s}{ds'
\frac{\delta^{2}}{\delta v_{\mu}^{2}(s')}}} \cdot \left[ m - \gamma_{\mu} \, \frac{\delta}{\delta v_{\mu}(s)} \right] } \, \delta( x -y + \int_{0}^{s}{ds' \ v(s')}) \\ \nonumber & &
\times \left.  \left(\exp{\left\{ -ig \int_{0}^{s}{ds' \left[v_{\mu}(s') \, A_{\mu}^{a}(y-\int_{0}^{s'}{v}) \tau^{a} + i \sigma_{\mu \nu} \, \mathbf{F}_{\mu \nu}^{a}(y-\int_{0}^{s'}{v}) \tau^{a} \right] }\right\}} \right)_{+}  \right|_{v_{\mu} \rightarrow 0}.
\end{eqnarray}

\noindent Then, one can insert a functional `resolution of unity' of form
\begin{equation}\label{Eq:A3}
1 = \int{\mathrm{d}[u] \, \delta[u(s') - \int_{0}^{s'}{ds'' \ v(s'')}]},
\end{equation}

\noindent and replace the delta-functional $\delta[u(s') - \int_{0}^{s'}{ds'' \ v(s'')}]$ with a functional integral over $\Omega$, and then the Green's function becomes~\cite{YMS2008}
\begin{eqnarray}\label{Eq:A4}
&& \mathbf{G}_{\mathrm{c}}(x,y|A) \\ \nonumber &=&  i \int_{0}^{\infty}{ds \ e^{-is m^{2}}} \, e^{- \frac{1}{2} \Tr{\ln{\left( 2h \right)}} } \, \int{d[u]} \, e^{ \frac{i}{4} \int_{0}^{s}{ds' \, [u'(s')]^{2} } } \, \delta^{(4)}(x - y + u(s)) \\ \nonumber & & \quad \times {\left[ m + i g \gamma_{\mu} A_{\mu}^{a}(y-u(s)) \tau^{a} \right]} \, \left( e^{ -ig \int_{0}^{s}{ds' \, u'_{\mu}(s') \, A_{\mu}^{a}(y-u(s')) \, \tau^{a}} + g \int_{0}^{s}{ds' \sigma_{\mu \nu} \, \mathbf{F}_{\mu \nu}^{a}(y-u(s')) \, \tau^{a}}} \right)_{+},
\end{eqnarray}

\noindent where $h(s_{1},s_{2})=\int_{0}^{s}{ds' \, \Theta(s_{1} - s') \Theta(s_{2} - s')}$.  To remove the $A$-dependence out of the linear (mass) term, one can replace $i g A_{\mu}^{a}(y-u(s)) \tau^{a}$ with $- \frac{\delta}{\delta u'_{\mu}(s)}$ operating on the ordered exponential so that
\begin{eqnarray}\label{Eq:A5}
&& \mathbf{G}_{\mathrm{c}}(x,y|A) \\ \nonumber &=&  i \int_{0}^{\infty}{ds \ e^{-is m^{2}}} \, e^{- \frac{1}{2} \Tr{\ln{\left( 2h \right)}} } \, \int{d[u]} \, e^{ \frac{i}{4} \int_{0}^{s}{ds' \, [u'(s')]^{2} } } \, \delta^{(4)}(x - y + u(s)) \\ \nonumber & & \quad \times {\left[ m - \gamma_{\mu} \frac{\delta}{\delta u'_{\mu}(s)} \right]} \, \left( e^{ -ig \int_{0}^{s}{ds' \, u'_{\mu}(s') \, A_{\mu}^{a}(y-u(s')) \, \tau^{a}} + g \int_{0}^{s}{ds' \sigma_{\mu \nu} \, \mathbf{F}_{\mu \nu}^{a}(y-u(s')) \, \tau^{a}}} \right)_{+}.
\end{eqnarray}

\noindent To extract the $A$-dependence out of the ordered exponential, one may use the following identities,
\begin{eqnarray}
1 &=& \int{d[\alpha] \, \delta{\left[ \alpha^{a}(s') + g u'_{\mu}(s') \, A^{a}_{\mu}(y-u(s'))\right]}},  \\ \nonumber 1 &=& \int{d[\mathbf{\Xi}] \, \delta{\left[ \mathbf{\Xi}^{a}_{\mu \nu}(s') - g \mathbf{F}_{\mu \nu}^{a}(y-u(s'))\right]} },
\end{eqnarray}

\noindent and the ordered exponential becomes
\begin{eqnarray}
& & \left( e^{ -ig \int_{0}^{s}{ds' \, u'_{\mu}(s') \, A_{\mu}^{a}(y-u(s')) \, \tau^{a}} + g \int_{0}^{s}{ds' \sigma_{\mu \nu} \, \mathbf{F}_{\mu \nu}^{a}(y-u(s')) \, \tau^{a}}} \right)_{+} \\ \nonumber &=& \mathcal{N}_{\Omega} \, \mathcal{N}_{\Phi} \, \int{d[\alpha] \, \int{d[\mathbf{\Xi}] \, \int{d[\Omega] \, \int{d[\mathbf{\Phi}] \, \left( e^{ i \int_{0}^{s}{ds' \, \left[ \alpha^{a}(s') - i \sigma_{\mu \nu} \, \mathbf{\Xi}_{\mu \nu}^{a}(s') \right] \, \tau^{a}}} \right)_{+} }}}} \\ \nonumber & & \quad \times e^{-i \int{ds' \, \Omega^{a}(s') \, \alpha^{a}(s')}  - i \int{ds' \, \mathbf{\Phi}^{a}_{\mu \nu}(s') \,  \mathbf{\Xi}^{a}_{\mu \nu}(s') } } \\ \nonumber & & \quad \times e^{- i g \int{ds' \, u'_{\mu}(s') \, \Omega^{a}(s') \, A^{a}_{\mu}(y-u(s')) } + i g \int{ds' \, \mathbf{\Phi}^{a}_{\mu \nu}(s') \, \mathbf{F}_{\mu \nu}^{a}(y-u(s'))}  },
\end{eqnarray}

\noindent where $\mathcal{N}_{\Omega}$ and $\mathcal{N}_{\Phi}$ are constants that normalize the functional representations of the delta-functionals.  All $A$-dependence is removed from the ordered exponential and the resulting form of the Green's function is exact (it entails no approximation).  Alternatively, extracting the $A$-dependence out of the ordered exponential can also be achieved by using the functional translation operator, and one writes
\begin{equation}
\left( e^{ + g \int_{0}^{s}{ds' \, \left[ \sigma_{\mu \nu} \, \mathbf{F}_{\mu \nu}^{a}(y-u(s')) \tau^{a} \right]} }\right)_{+}  = \left. e^{g \int_{0}^{s}{ds' \, \mathbf{F}_{\mu \nu}^{a}(y-u(s')) \, \frac{\delta}{\delta \mathbf{\Xi}_{\mu\nu}^{a}(s')}}} \cdot \left( e^{\int_{0}^{s}{ds' \, \left[ \sigma_{\mu \nu} \, \mathbf{\Xi}_{\mu \nu}^{a}(s') \tau^{a} \right]} }\right)_{+} \right|_{\mathbf{\Xi} \rightarrow 0}.
\end{equation}

For the closed-fermion-loop functional $\mathbf{L}[A]$, one can write~\cite{HMF2}
\begin{equation}\label{Eq:ClosedFermionLoopFunctional02}
\mathbf{L}[A] = - \frac{1}{2} \, \int_{0}^{\infty}{\frac{ds}{s} \, e^{-ism^{2}} \, \left\{\Tr{\left[ e^{-is(\gamma \cdot \Pi)^{2}} \right]} - \left\{ g=0 \right\} \right\}},
\end{equation}

\noindent where the trace $\Tr{}$ sums over all degrees of freedom, space-time coordinates, spin and color.  The Fradkin representation proceeds along the same steps as in the case of $\mathbf{G}_{\mathrm{c}}[A] $, and the closed-fermion-loop functional reads
\begin{eqnarray}\label{Eq:LFradkin01}
\mathbf{L}[A] &=&  - \frac{1}{2} \int_{0}^{\infty}{\frac{ds}{s} \, e^{-is m^{2}}} \, e^{- \frac{1}{2} \Tr{\ln{(2h)}} } \\ \nonumber && \quad \times \int{d[v]} \, \delta^{(4)}(v(s)) \, e^{ \frac{i}{4} \int_{0}^{s}{ds' \, [v'(s')]^{2} } } \\ \nonumber & & \quad \times \int{d^{4}x \, \tr{\left( e^{ -ig \int_{0}^{s}{ds' \, v'_{\mu}(s') \, A_{\mu}^{a}(x-v(s')) \, \tau^{a}} + g \int_{0}^{s}{ds' \sigma_{\mu \nu} \, \mathbf{F}_{\mu \nu}^{a}(x-v(s')) \, \tau^{a}}} \right)_{+}} } \\ \nonumber & & - \left\{ g = 0 \right\},
\end{eqnarray}

\noindent where the trace $\tr{}$ sums over color and spinor indices.  Also, Fradkin's variables have been denoted by $v(s')$, instead of $u(s')$, in order to distinguish them from those appearing in the Green's function $\mathbf{G}_{\mathrm{c}}[A] $. One finds
\begin{eqnarray}\label{Eq:LFradkin02}
\mathbf{L}[A] &=&  - \frac{1}{2} \int_{0}^{\infty}{\frac{ds}{s} \, e^{-is m^{2}}} \, e^{- \frac{1}{2} \Tr{\ln{(2h)}} } \\ \nonumber && \quad \times \mathcal{N}_{\Omega} \, \mathcal{N}_{\Phi} \int{d^{4}x \, \int{\mathrm{d}[\alpha] \, \int{\mathrm{d}[\Omega] \, \int{\mathrm{d}[\mathbf{\Xi}] \, \int{\mathrm{d}[\mathbf{\Phi}] \, }}}}}  \\ \nonumber & & \quad \times \int{d[v] \, \delta^{(4)}(v(s)) \, e^{ \frac{i}{4} \int_{0}^{s}{ds' \, [v'(s')]^{2} } }  } \\ \nonumber & & \quad \times \ e^{-i \int{ds' \, \Omega^{a}(s') \, \alpha^{a}(s')} - i \int{ds' \, \mathbf{\Phi}^{a}_{\mu \nu}(s') \,  \mathbf{\Xi}^{a}_{\mu \nu}(s') } }  \cdot  \tr{\left( e^{ i \int_{0}^{s}{ds' \, \left[ \alpha^{a}(s') - i \sigma_{\mu \nu} \, \mathbf{\Xi}_{\mu \nu}^{a}(s') \right] \, \tau^{a}}} \right)_{+}} \\ \nonumber & & \quad \times  e^{- i g \int_{0}^{s}{ds' \, v'_{\mu}(s') \, \Omega^{a}(s') \, A^{a}_{\mu}(x - v(s'))} - 2 i g \int{d^{4}z \, \left(\partial_{\nu}  \mathbf{\Phi}^{a}_{\nu \mu}(z) \right) \, A^{a}_{\mu}(z) }} \\ \nonumber & & \quad \times e^{ + i g^{2} \int{ds' \, f^{abc} \mathbf{\Phi}^{a}_{\mu \nu}(s') \, A^{b}_{\mu}(x- v(s')) \, A^{c}_{\nu}(x- v(s')) }}  \\ \nonumber & & - \left\{ g = 0 \right\},
\end{eqnarray}

\noindent where the same properties as those of $\mathbf{G}_{\mathrm{c}}[A] $ can be read off readily.

\section{\label{AppB} Effective Locality versus Transverse Average}
Before transverse imprecision was introduced, EL had the effect of attaching to the representative symbol $[f \cdot \chi(w)]^{-1}$ of each gluon bundle, exchanged between quark and/or antiquark of respective CM coordinates $y_{1}$ and $y_{2}$, a pair of delta functions, $\delta^{(4)}(w - y_{1} + u(s_{1})) \delta^{(4)}(y_{1} - y_{2} + \bar{u}(s_{2}) - u(s_{1}))$, as used in the text, or the pair $\delta^{(4)}(w - y_{1} + s_{1} p_{1}) \delta^{(4)}(y_{1} - y_{2} + s_{2} p_{2} - s_{1} p_{2})$ as used in an eikonal approximation~\cite{Fried2009_QCD1}.  For either case one finds fixed values of $w_{0}$ and $w_{\mathrm{L}}$, and $\vec{w}_{\perp} = \vec{y}_{1\perp} =- \vec{y}_{2\perp}$.  Then, as claimed in the text, the Halpern FI can be reduced to an ordinary set of $\int{\mathrm{d}^{n}\chi}$ integrals. In the process, though, one makes a systematic error, of the eikonal-type, by neglecting variations of the impact parameter or, correspondingly, of momentum transfer in the core parts of the matrix element.  In the context of the exact expression of the first pair of delta functions above, that \emph{ad hoc} approximation avoided the much more complicated analysis of the transverse Fradkin's difference $u_{\perp}(s_{1}) - \bar{u}_{\perp}(s_{2})$.

With transverse imprecision now being included, the situation changes for the better, in the sense that no such approximation need be made.  But this change now requires a slightly more complicated justification of the argument which replaces Halpern's FI by a set of ordinary integrals.  For the question arises if this useful simplification is also true when the $\vec{w}_{\perp}$ inside the $[f \cdot \chi(w)]^{-1}$ factor is itself given by $\vec{y}^{\, \prime}_{\perp}$, and is being integrated over the $\int{\mathrm{d}^{2} \vec{y}^{\, \prime}}$ in that exponential factor, as in the discussion of the text leading to (\ref{Eq:46}).  It was there noted that the replacement of that $\vec{w}_{\perp}$ by $\vec{y}_{1\perp}$ or $-\vec{y}_{2\perp}$ is a reasonable approximation.

The following argument is intended to give that approximation a more detailed justification.

Consider the Halpern FI
\begin{eqnarray}\label{Eq:B1}
\mathcal{N} \int{\mathrm{d}[\chi] \, [\det{(f \cdot \chi)}]^{-\frac{1}{2}} \, e^{\frac{i}{4}\int{\mathrm{d}^{4}w \, \chi^{2}} + ig\int{\mathrm{d}^{2}\vec{y}^{\, \prime}_{\perp} \, \mathfrak{a}(\vec{y}_{1\perp} - \vec{y}^{\, \prime}_{\perp}) \, \mathfrak{a}(\vec{y}_{2\perp} - \vec{y}^{\, \prime}_{\perp}) [f \cdot \chi(\vec{y}^{\, \prime}_{\perp})]^{-1}} } },
\end{eqnarray}
											
\noindent where $y'_{\mu} = (y_{0}; y_{\mathrm{L}}, \vec{y}^{\, \prime}_{\perp})$, the normalization is defined so that the FI of (\ref{Eq:B1}) equals  1 when $g = 0$. The dependence of color, time and longitudinal coordinate has been omitted for simplification of presentation.

As in the definition of this or any such FI, $\int{\mathrm{d}^{4}w \, \chi^{2} }$ is understood to mean $\delta^{4} \, \sum_{\ell=1}^{N}{\chi^{2}_{\ell}}$, where the subscript $\ell$ denotes the value of $\chi$ at the space-time point $w_{\perp \ell}$, and $\delta^{4}$ corresponds to a small volume surrounding that point, which is to become arbitrarily small as $N$ becomes arbitrarily large~\cite{Zee:2010}.  As mentioned in the text by the end of Sec.~\ref{SEC5}, residual $\delta$-dependence will be re-expressed in terms of physically significant quantities as a last step; but for the following argument, all the transverse coordinate differences are to be taken as arbitrarily small.

Now, re-scale the $\chi_{\ell}$ variables such that $\delta^{2} \chi_{\ell} = \bar{\chi}_{\ell}$, and re-write (\ref{Eq:B1}) as
\begin{eqnarray}\label{Eq:B2}
\bar{\mathcal{N}} \int{\mathrm{d}[\bar{\chi}] \, [\det{(f \cdot \bar{\chi})}]^{-\frac{1}{2}} \, e^{\frac{i}{4} \, \sum_{\ell}{\bar{\chi}_{\ell}^{2}} + ig \delta^{2} \, \int{\mathrm{d}^{2}\vec{y}^{\, \prime}_{\perp} \, \mathfrak{a}(\vec{y}_{1\perp} - \vec{y}^{\, \prime}_{\perp}) \, \mathfrak{a}(\vec{y}_{2\perp} - \vec{y}^{\, \prime}_{\perp}) [f \cdot \bar{\chi}(\vec{y}^{\, \prime}_{\perp})]^{-1}} } }.
\end{eqnarray}

\noindent Let us also break up the $\int{\mathrm{d}^{2}\vec{y}^{\, \prime}_{\perp}}$ integral into an infinite series of terms: one is free to choose the individual $\vec{y}^{\, \prime}_{\perp}$ coordinates as exactly those which define the transverse positions of the $\bar{\chi}_{\ell} = \bar{\chi}(w_{\ell})$.  In this way, (\ref{Eq:B2}) may be re-written as
\begin{eqnarray}\label{Eq:B3}
\bar{\mathcal{N}} \int{\mathrm{d}[\bar{\chi}] \, [\det{(f \cdot \bar{\chi})}]^{-\frac{1}{2}} \, e^{\frac{i}{4} \, \sum_{\ell}{\bar{\chi}_{\ell}^{2}} + ig \delta^{2} \, \Delta^{2}_{y_{\perp}} \, \sum_{\ell}{\mathfrak{a}(\vec{y}_{1\perp} - \vec{y}^{\, \prime}_{\perp\ell}) \, \mathfrak{a}(\vec{y}_{2\perp} - \vec{y}^{\, \prime}_{\perp \ell}) [f \cdot \bar{\chi}(\vec{y}^{\, \prime}_{\perp\ell})]^{-1}} } },
\end{eqnarray}

\noindent where $\Delta^{2}_{y_{\perp}}$ is understood as a true infinitesimal quantity, and where, for simplicity, we suppress explicit dependence on $y_{0}$ and $y_{\mathrm{L}}$.  But now (\ref{Eq:B3}) may be written as the product of $N$ integrals,
\begin{eqnarray}\label{Eq:B4}
&& \prod_{\ell}^N \bar{\mathcal{N}}_{\ell} \int{\mathrm{d}^{n}\bar{\chi}(\vec{y}^{\, \prime}_{\perp \ell}) \, [\det{(f \cdot \bar{\chi}(\vec{y}^{\, \prime}_{\perp \ell}))}]^{-\frac{1}{2}} \, e^{\frac{i}{4} \, {\bar{\chi}^{2}(\vec{y}^{\, \prime}_{\perp \ell})} + ig \delta^{2} \, \Delta^{2}_{y_{\perp}} \, {\mathfrak{a}(\vec{y}_{1\perp} - \vec{y}^{\, \prime}_{\perp\ell}) \, \mathfrak{a}(\vec{y}_{2\perp} - \vec{y}^{\, \prime}_{\perp \ell}) [f \cdot \bar{\chi}(\vec{y}^{\, \prime}_{\perp\ell})]^{-1}} } } \\ \nonumber &\equiv& \prod_{\ell}^N{\mathbb{F}(ig \delta^{2} \, \Delta^{2}_{y_{\perp}} \, \mathfrak{a}(\vec{y}_{1\perp} - \vec{y}^{\, \prime}_{\perp\ell}) \, \mathfrak{a}(\vec{y}_{2\perp} - \vec{y}^{\, \prime}_{\perp \ell}))} \\ \nonumber &\equiv& \prod_{\ell}^N{\mathbb{F}(z_{\ell})},
\end{eqnarray}
												
\noindent where (\ref{Eq:B4}) denotes the normalized product of all such $(\vec{y}^{\, \prime}_{\perp \ell})$-valued integrals, and $\mathbb{F}(z_{\ell})$ denotes the ordinary integral $\int{\mathrm{d}^{n}\bar{\chi}_{\ell}}$ over the variable associated with $\vec{y}^{\, \prime}_{\perp \ell}$.  That integral is well defined in the sense that, for $|z_{\ell}| < 1$, as is the case here, it can be expressed as an absolutely-convergent series, or as a converging integral over a set of eigenvalues in a random matrix calculation.

One then expects to be able to write $\mathbb{F}(z)$ in terms of its Fourier transform,
\begin{equation}
\mathbb{F}(z_{\ell}) = \int{\mathrm{d}\varrho \, \tilde{\mathbb{F}}(\varrho) \, e^{i z_{\ell} \varrho}},
\end{equation}

\noindent where the normalization condition of (\ref{Eq:B4}) stipulates that $\int{\mathrm{d}\varrho \, \tilde{\mathbb{F}}(\varrho)} = 1$.  Since $z_{\ell}$ is proportional to the infinitesimal $\Delta^{2}_{y_{\perp}}$, one may expand in powers of $z_{\ell}$,
\begin{equation}
\mathbb{F}(z_{\ell}) = \int{\mathrm{d}\varrho \, \tilde{\mathbb{F}}(\varrho) \, \left[ 1 + i z_{\ell} \varrho + \cdots \right]} = 1 + i \int{\mathrm{d}\varrho \, \varrho \, \tilde{\mathbb{F}}(\varrho)} \, z_{\ell} + \cdots,
\end{equation}

\noindent so that (\ref{Eq:B4}) becomes approximately
\begin{eqnarray}\label{Eq:B5}
&& \prod_{\ell}{ \left[ 1 + i \int{\mathrm{d}\varrho \, \varrho \, \tilde{\mathbb{F}}(\varrho)} \, z_{\ell} \right]} \\ \nonumber &=& 1 + i \int{\mathrm{d}\varrho \, \varrho \, \tilde{\mathbb{F}}(\varrho)} \, \sum_{\ell}{z_{\ell}} \\ \nonumber &=& 1 + i \int{\mathrm{d}\varrho \, \varrho \, \tilde{\mathbb{F}}(\varrho)} \cdot ig  \delta^{2} \, \int{\mathrm{d}^{2}\vec{y}^{\, \prime}_{\perp} \, \mathfrak{a}(\vec{y}_{1\perp} - \vec{y}^{\, \prime}_{\perp}) \, \mathfrak{a}(\vec{y}_{2\perp} - \vec{y}^{\, \prime}_{\perp})}.
\end{eqnarray}

\noindent With
\begin{equation}
\int{\mathrm{d}^{2}\vec{y}^{\, \prime}_{\perp} \, \mathfrak{a}(\vec{y}_{1\perp} - \vec{y}^{\, \prime}_{\perp}) \, \mathfrak{a}(\vec{y}_{2\perp} - \vec{y}^{\, \prime}_{\perp})} = \int{\frac{\mathrm{d}^{2}q}{(2 \pi)^{2}}} \, \left| \tilde{\mathfrak{a}}(q) \right|^{2} \, e^{iq \cdot (\vec{y}_{1\perp} - \vec{y}_{2\perp})} \equiv \varphi(\vec{b}),
\end{equation}

\noindent Eq.~(\ref{Eq:B4}) becomes
\begin{equation}\label{Eq:B9}
\int{\mathrm{d}\varrho \, \tilde{\mathbb{F}}(\varrho) \, \left[ 1 + i \varrho (ig \delta^{2}) \varphi(\vec{b}) \right] },
\end{equation}

\noindent which is just the first-order expansion of the result obtained in the text when $\vec{y}^{\, \prime}_{\perp}$ was shifted to $\vec{y}^{\, \prime}_{1\perp}$ or $-\vec{y}^{\, \prime}_{2\perp}$.  And since $g \, \delta^{2} \, \varphi$ is expected to be small, $\delta^{2} \, \varphi \ll 1$, it is, in effect, equivalent to
\begin{equation*}
\int{\mathrm{d}\varrho \, \tilde{\mathbb{F}}(\varrho) \, e^{i \varrho (ig \delta^{2}) \varphi(\vec{b})} },
\end{equation*}

\noindent which is just the integral of (\ref{Eq:B2}) when the intuitively equivalent change $\vec{y}^{\, \prime}_{\perp} \rightarrow \vec{y}^{\, \prime}_{1\perp}$ has been made in the argument of $(f \cdot \bar{\chi})^{-1}$, and after the residual $\delta^{2}$ dependence has been continued to the measurably-significant value of $1/(\mu E)$. Only the first-order form, corresponding to (\ref{Eq:B9}) is needed in the calculations of quark- and nucleon-binding.

%
%



\begin{thebibliography}{}
%
%
\bibitem[Reinhardt(1993)]{Reinhardt:1993}
H. Reinhardt, K. Langfeld, and L. v. Smekal, \emph{Phys. Lett.} B\textbf{300}, 111 (1993).

\bibitem{Ralf:2008}
R. Hofmann, Int. J. Mod. Phys. A{\bf{20}}, 4123 (2006); Erratum-ibid. A{\bf{21}} 6515 (2006).

\bibitem{DanielDF:2011}
D. D. Ferrante, G. S. Guralnik, Z. Guralnik and C. Pehlevan, {\it{Complex Path Integrals and the Space of Theories}}, Brown Univsersity Preprint: BROWN-HET-1611 (2011).

\bibitem{Fried2009_QCD1}
H. M. Fried, Y. Gabellini, T. Grandou and Y.-M. Sheu, Eur. Phys. J. C\textbf{65}, 395 (2010).

\bibitem{Fradkin1966}
E. S. Fradkin, Nucl. Phys. \textbf{76}, 588 (1966).

\bibitem{HMF2}
H. M. Fried, \textit{Basics of Functional Methods and Eikonal Models} (Editions Fronti\`{e}res, Gif-sur-Yvette Cedex, France 1990)

\bibitem{HMF1}
H. M. Fried, \textit{Functional Methods and Models in Quantum Field Theory} (The MIT Press, Cambridge, MA 1972)

\bibitem{bctg:2012}
B. Candelpergher and T. Grandou, work in progress.

\bibitem{Halpern1977a}
M. B. Halpern, Phys. Rev. D\textbf{16}, 1798 (1977).

\bibitem{Halpern1977b}
M. B. Halpern, Phys. Rev. D\textbf{16}, 3515 (1977).

\bibitem {Zee:2010}
A. Zee, \textit{Quantum Field Theory in a Nutshell} (Princeton University Press, Princeton 2010)

\bibitem{Guay:2004}
A. Guay, {\it{Geometrical apects of local gauge symmetry}}, (2004), \url{http://philsci-archive.pitt.edu/id/eprint/2133}.

\bibitem{Huang:1977}
K. Huang and D. R. Stump, Phys. Rev. Lett. \textbf{37}, 545 (1976); Phys. Rev. D\textbf{15}, 3660 (1977).

\bibitem{Fried5:2012}
H. M. Fried, Y. Gabellini, T. Grandou and Y.-M. Sheu, arXiv:1203.6137v1 [hep-ph] (2012).

\bibitem{hftg:2012}
H. M. Fried, T. Grandou and Y.-M. Sheu, work in completion.

\bibitem{11tg:2011}
T. Grandou, {\it{On Some Aspects of the QCD Effective Locality}}, Proceedings of the Eleventh Workshop on Non-Perturbative QCD, Paris, June 2011, Edited by B. M\"uller and C. I. Tan, \url{http://www.slac.stanford.edu/econf/C1106064/}.

\bibitem{Ref-Machine}
See, for example, the seminal papers by Y. Nambu, Phys. Lett. B\textbf{80}, 372 (1979); M. Luscher, K. Symanzik, P. Weisz, Nucl. Phys. B\textbf{173}, 365 (1980); M. Luscher, Nucl. Phys. B\textbf{180}, 317 (1981).

\bibitem{Quantics JMLL}
F. Balibar, A. Laverne and J. M. Levy Leblond,{\it{ Quantique:El\'ements}}, \url{http://cel.archives-ouvertes.fr/docs/00/13/61/89/PDF/elem_5fev07.pdf}.

\bibitem{YMS2008}
Y.-M. Sheu, {\it{Finite-Temperature Quantum Electrodynamics: General Theory and Bloch-Nordsieck Estimates of Fermion Damping in a Hot Medium}}, PhD Thesis, Brown University, May 2008.


\bibitem{Fried2000}
H. M. Fried, Y. Gabellini and J. Avan, Eur. Phys. J. C\textbf{13}, 699 (2000).

\bibitem{levy}
For example, see \url{http://en.wikipedia.org/wiki/Stable_distribution}.


\bibitem{HMF3}
H. M. Fried, \textit{Green's Functions and Ordered Exponentials} (Cambridge University Press, Cambridge 2002)

\bibitem{Lapidus2000}
G. W. Johnson and M. L. Lapidus, \textit{The Feynman Integral and Feynman's Operational Calculus} (Oxford University Press, Oxford 2000)

\bibitem{GR}
I. S. Gradshteyn and I. M. Ryzhik, \textit{Table of Integrals, Series and Products} (Academic Press, London 1994) formula 8.253.1

\bibitem{Schiff}
L. I. Schiff, \textit{Quantum Mechanics, 3rd Ed.} (McGraw-Hill, 1968)

\bibitem{Note}
$w_{0}= (0_{0},\vec{y}_{\perp}, 0_{\mathrm{L}})$, correcting the expression given in Ref.~\cite{Fried2009_QCD1}

\bibitem{Fried1983}
H.M. Fried, K. Kang and B. H. J. McKellar, Phys. Rev. A{\textbf{28}}, 738 (1983).


\end{thebibliography}


\end{document}